\documentclass[10pt,conference]{IEEEtran}
\usepackage{authblk}
\usepackage{amsmath,amsfonts}
\usepackage{algorithmic}
\usepackage{algorithm}
\usepackage{array}
\usepackage[caption=false,font=normalsize,labelfont=sf,textfont=sf]{subfig}
\usepackage{textcomp}
\usepackage{stfloats}
\usepackage{url}
\usepackage{verbatim}
\usepackage{graphicx}
\usepackage{cite}
\usepackage{xcolor}
\usepackage{soul}
\usepackage{amssymb}
\usepackage{enumitem}
\usepackage{booktabs}

\usepackage{subfig}
\usepackage{arydshln}
\usepackage{fancyhdr}
\usepackage{varwidth}
\usepackage{arydshln}
\usepackage{multirow}
\usepackage{mathtools}

\usepackage{datetime}
\usepackage{kotex}
\usepackage{soul}
\usepackage{pifont}
\usepackage{colortbl}
\usepackage{tabularray}

\newcommand{\sys}{\textsf{\small CF-DETR}}
\newcommand{\npfpstar}{\mbox{NPFP$^{**}$}}
\newcommand{\npfpc}{\mbox{NPFP$^\textsf{C}$}}
\newcommand{\npfpcfb}{\mbox{NPFP$^\textsf{C[F]}$}}
\newcommand{\npfpcbf}{\mbox{NPFP$^\textsf{[C]F}$}}
\newcommand{\npfpcbfb}{\mbox{NPFP$^\textsf{[C][F]}$}}

\newcommand{\jl}[1]{{\color{black}{#1}}} %red
\newcommand{\jlcmt}[1]{{\color{blue}{[JL] #1}}} %blue

\newcommand{\hb}[1]{{\color{black}{#1}}}          %magenta
\newcommand{\hbcmt}[1]{{\color{cyan}{[HB: #1]}}}    %cyan
\definecolor{caribbeangreen}{rgb}{0.0, 0.6, 0.6}
\newcommand{\kdh}[1]{{\color{orange}{#1}}}
\newcommand{\dhcmt}[1]{{\color{caribbeangreen}{[DH: #1]}}}

\newcommand{\remove}[1]{}

\newtheorem{lemma}{Lemma}
\newtheorem{theorem}{Theorem}

\begin{document}

%\title{MC-MOT: Real-Time DNN-based Criticality-Aware 
%%Mixed-Critical 
%Multi-Object Tracking}
\title{CF-DETR: Coarse-to-Fine Transformer for Real-Time Object Detection} 
%\jl{\st{DNN-based}} \jl{\st{Multi-Camera}} Multi-Object Tracking}
%\title{A Sample Article Using IEEEtran.cls\\ for IEEE Journals and Transactions}

\author{
Woojin Shin$^{1}$, Donghwa Kang$^{2}$, Byeongyun Park$^{1}$, Brent Byunghoon Kang$^{2}$, \\
Jinkyu Lee$^{3}$ and Hyeongboo Baek$^{1}$ 
\\
\\
$^{1}$University of Seoul \quad
$^{2}$KAIST \quad
$^{3}$Sungkyunkwan University
}

\markboth{Journal of \LaTeX\ Class Files,~Vol.~14, No.~8, April~2022}%
{Shell \MakeLowercase{\textit{et al.}}: A Sample Article Using IEEEtran.cls for IEEE Journals}

\IEEEpubid{0000--0000/00\$00.00~\copyright~2021 IEEE}

\maketitle

\thispagestyle{plain}
\pagestyle{plain}

\begin{abstract}
Detection Transformers (DETR) are increasingly adopted in autonomous vehicle (AV) perception systems due to their superior accuracy over convolutional networks. However, concurrently executing multiple DETR tasks presents significant challenges in meeting firm real-time deadlines (R1) and high accuracy requirements (R2), particularly for safety-critical objects, while navigating the inherent latency-accuracy trade-off under resource constraints. 
Existing real-time DNN scheduling approaches often treat models generically, failing to leverage Transformer-specific properties for efficient resource allocation. 
To address these challenges, we propose \sys{}, an integrated system featuring a novel coarse-to-fine Transformer architecture and a dedicated real-time scheduling framework \npfpstar{}. 
\sys{} employs three key strategies (A1: coarse-to-fine inference, A2: selective fine inference, A3: \hb{multi-level batch inference}) that exploit Transformer properties to dynamically adjust patch granularity and attention scope based on object criticality, aiming to satisfy R2. 
\hb{The \npfpstar{} scheduling framework (A4) orchestrates these adaptive mechanisms A1--A3. 
It partitions each DETR task into a safety-critical coarse subtask for guaranteed critical object detection within its deadline (ensuring R1), and an optional fine subtask for enhanced overall accuracy (R2), while managing individual and batched execution.}
Our extensive evaluations on server, GPU-enabled embedded platforms, and actual AV platforms demonstrate that \sys{}, under an \npfpstar{} policy, successfully meets strict timing guarantees for critical operations and achieves significantly higher overall and critical object detection accuracy compared to existing baselines across diverse AV workloads.
\end{abstract}

%\begin{IEEEkeywords}
	%Multi-object tracking, real-time scheduling, timing guarantee
%\end{IEEEkeywords}

\section{Introduction}
\label{sec:introduction}

Detection Transformers (DETR) have significantly advanced object detection, outperforming traditional convolutional networks in accuracy and, in some instances, inference speed, making them highly attractive for autonomous vehicle (AV) perception systems~\cite{ZLX24, ZSL20}. 
This superior performance stems from their global \textit{attention} mechanism, which processes images as split \textit{patch} collections to capture comprehensive contextual relationships~\cite{CMS20}. 
However, this powerful mechanism also incurs substantial computational costs that scale markedly with input patch count, creating a pronounced latency-accuracy trade-off~\cite{YVA22, BFD22}. 
For AVs concurrently running multiple DETR tasks (e.g., per camera) under stringent real-time deadlines (R1) and high accuracy demands (R2)---especially for safety-critical objects---this trade-off presents a formidable challenge~\cite{KLC22, KCK22}. 
The fundamental design goal is thus to satisfy both R1 and R2, prioritizing critical objects within typical AV resource constraints~\cite{KCK22, KLB25, SXM22}.

Since the advent of vision Transformers (ViTs) approximately five years ago, various lightweighting techniques have been explored. 
However, adaptively managing the computational load of such models specifically for multi-task DETR execution within safety-critical real-time systems remains a significant, largely unaddressed challenge.
Prevailing real-time DNN scheduling studies~\cite{KCK22, ZBL18, XYK19, KLC22, JJK22} predominantly focus on convolutional architectures or treat DNNs as monolithic black boxes, lacking a nuanced understanding of internal Transformer properties like the critical impact of patch granularity and attention scope on the latency-accuracy trade-off. 
Among recent lightweight ViT methods~\cite{YVA22, BFD22}, coarse-to-fine (CF) strategies are gaining traction in classification (e.g., CF-ViT~\cite{CLL23}); however, their systematic application to \textit{detection} transformers is a largely unaddressed area.
This is primarily due to detection's sophisticated spatial understanding requirements and the inherent challenge of integrating fine-grained adaptability with real-time scheduling. 
Our key insight, drawn from experimental observations (Sec.~\ref{subsec:observation}), is that a CF approach tailored for DETR (\textit{first} proposed in this paper) not only effectively navigates the latency-accuracy trade-off but also excels in prioritizing responsiveness and accuracy for safety-critical objects under constrained resources.

In this paper, we propose \sys{}, an integrated system featuring a novel coarse-to-fine DETR architecture coupled with a dedicated real-time scheduling framework. 
\jl{The \sys{} architecture}
%\sys{} (the architecture) 
primarily tackles the challenge by leveraging fundamental Transformer properties to navigate the latency-accuracy trade-off while considering object criticality. \sys{} proposes three complementary approaches A1–A3 (to be presented) that optimize DETR’s costly self-attention stage. As the computational complexity of self-attention scales with the number of patches, there is an explicit trade-off between fewer patches, resulting in faster inference but potentially detecting only larger critical objects, and more patches, yielding slower but more accurate detection for objects of all sizes. \sys{}
strategically leverages this property by dynamically adjusting patch granularity and attention scope based on each input’s object criticality.

% \noindent A1. \textit{Coarse-to-fine inference (image-level allocation)}: 
% Some images are \textit{easy} to achieve high accuracy on, achieving sufficient accuracy even with a coarse patch division (e.g., $3\times3$), typically containing only large, critical objects. 
% Conversely, others are \textit{hard}, containing objects of varied sizes, which require finer patch divisions (e.g., $9\times9$) for accurate detection.
% In A1, \sys{} first processes each image rapidly with a coarse-grained DETR inference. 
% It then identifies images exhibiting low-confidence detections (hard images) for further refinement at a finer granularity in A2. 

\begin{itemize}
\item [A1.] 
\textit{Coarse-to-fine inference} (image-level allocation): 
Some images are \textit{easy} to achieve high accuracy on, achieving sufficient accuracy even with a coarse patch division (e.g., $3\times3$), typically containing only large, critical objects. 
Conversely, others are \textit{hard}, containing objects of varied sizes, which require finer patch divisions (e.g., $9\times9$) for accurate detection.
In A1, \sys{} first processes each image rapidly with a coarse-grained DETR inference. 
It then identifies images exhibiting low-confidence detections (\textit{hard} images) for further refinement at a finer granularity in A2. 
\end{itemize}

\begin{itemize}
\item [A2.] 
\textit{Selective fine inference} (region-level allocation): Even within a hard image, not all regions are equally challenging—many parts containing large and critical objects of the image can be handled with coarse patches, while only certain regions need finer detail. 
Building on A1, A2 identifies and refines only the hard regions within hard images using more patches, thus avoiding full image reprocessing and dedicating computation efficiently to the areas that benefit most.
\end{itemize}

\begin{itemize}
\item [A3.] \hb{\textit{Multi-level batch inference} (batch-level allocation): Effective batching is crucial for throughput in AV systems running multiple concurrent DNN tasks, such as processing data from multiple cameras. A3 optimizes batch processing through a multi-level approach: for initial coarse-grained inference A1, \sys{} performs \textit{image-level} batch execution across multiple input frames. Subsequently, leveraging outputs from A1 and A2, it executes \textit{patch-level} batch inference exclusively on consolidated fine-grained hard regions from multiple tasks.}
\end{itemize}

To manage these adaptive \sys{} mechanisms A1--A3 across multiple DETR instances and effectively address the multi-task coordination challenge, we introduce (A4) the \npfpstar{} (Non-Preemptive Fixed-Priority with generic coarse-to-fine policy ``**'') scheduling framework. 
This framework orchestrates the \sys{} architecture by partitioning each DETR task into two key components: a safety-critical coarse subtask (primarily executing A1) and an optional, adaptive fine subtask (mainly executing A2). 
While the timely completion of coarse subtasks (without batching) is guaranteed by assigning higher priority to coarse subtasks over fine subtasks, the framework's flexibility allows different execution modes for coarse and fine subtasks, each with either individual or batched processing. 
To effectively integrate batching for improved overall accuracy without compromising timing guarantees, we develop novel batch assignments for each of the coarse and fine batches, which not only minimize the run-time overhead but also utilize the properties of batching and DETR.

% Paragraph 5: Experimental validation summary
Our extensive experimental evaluation validates the \sys{}'s effectiveness when orchestrated by an \npfpstar{} policy. Conducted on representative server and GPU-enabled embedded platforms, and demonstrated on an actual AV platform, \sys{} consistently achieved superior detection accuracy (overall and critical mAP) compared to baselines like standard DETR scheduling and state-of-the-art real-time DNN (RT-DNN) systems (e.g., DNN-SAM~\cite{KCK22}). 
This enhanced accuracy, particularly for vital critical detections, was attained with competitive or improved throughput (FPS), effectively managing the latency-accuracy trade-off while ensuring robust performance and precise localization crucial for AV safety.

% Paragraph 6: Contribution (non-itemized)
\textbf{Contribution.} 
Our approach fundamentally diverges from existing methods through its Transformer-specific co-design, without requiring additional sensors like LiDAR. 
Unlike generic RT-DNN systems (e.g., DNN-SAM), unaware of internal model characteristics (like attention mechanisms or patch granularity sensitivity) and thus allocating resources inefficiently for DETR models, our method leverages deep insights into these properties for effective optimization (A1 and A2). Furthermore, it incorporates Transformer-aware batching techniques (such as patch-level batch inference, A3) that yield significantly higher accuracy gains relative to computational cost compared to conventional batch-DNN methods ill-suited for complex Transformer workloads. 
This holistic, model-aware strategy addresses the limitations of applying generic real-time or batching solutions to multi-DETR systems' unique challenges.

% Paragraph 7: Contribution (itemized)
The key contributions of this paper are:
\begin{itemize}[leftmargin=*]
    \item We propose the \sys{} incorporating novel strategies A1--A3 exploiting Transformer properties (e.g., attention, patch granularity) to efficiently mitigate DETR's latency-accuracy trade-off.
    \item We introduce the \npfpstar{} scheduling framework A4, encompassing various instantiable policies, synergistically integrating A1--A3 strategies to achieve R1 and R2.
    \item We provide extensive empirical validation, including evaluations on an actual AV platform, demonstrating that the \sys{} significantly outperforms baselines in both overall detection accuracy (mAP) and real-time performance (FPS) across diverse AV workloads, ensuring critical object responsiveness and precise localization.
\end{itemize}

%\begin{figure}[t]
%    \centering
%    % \includegraphics[width=1\linewidth]{figures/detr.pdf}
%    \includegraphics[width=1\linewidth]{figures/02observation/DETR_overview.pdf}
%    \caption{...}
%    \label{fig:detr}
%\end{figure}

\section{Target system and observation}
This section explains our target system and presents key observations from a measurement-based case study that inform the design of \sys{}.

\subsection{Target system: detection Transformer}
\label{subsec:background}
\begin{figure}[t]
    \centering    
    \includegraphics[width=1\linewidth]{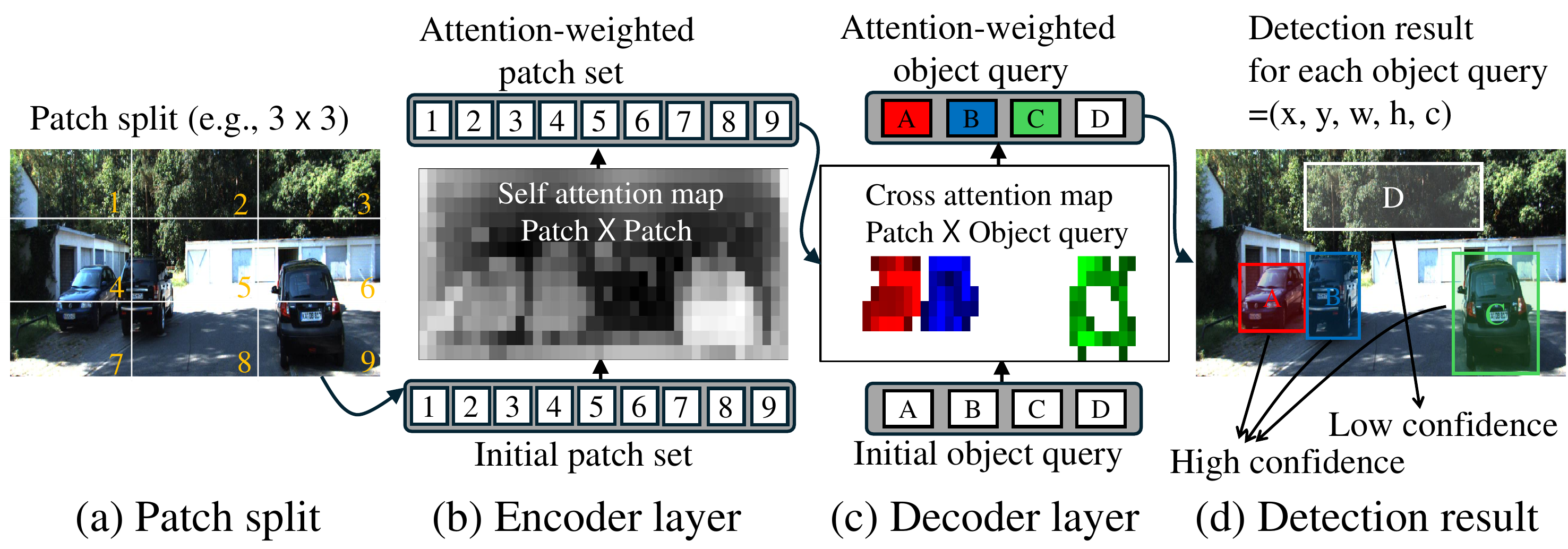}
    \caption{Overview of the DETR pipeline for object detection. The input image is divided into patches (a), which are encoded and processed via self-attention in the Transformer encoder (b). Learned object queries in the Transformer decoder interact with encoded patches through cross-attention (c), producing bounding boxes, class labels, and confidence scores (d).}
    \label{fig:background_detr}
\end{figure}

AVs typically rely on multiple cameras (e.g., front, side, rear) that each run a multi-object detection (MOD) model to spot vehicles, pedestrians, and other obstacles in every frame. 
Unlike traditional convolutional detectors (e.g., YOLO series~\cite{RDG16}) that only capture local features, DETR employs a patch-based representation and a Transformer architecture to model global context. 
The input image is first divided into a grid of small patches (e.g., a 3×3 grid in Fig.~\ref{fig:background_detr}(a)), and each patch is encoded as an embedding token (i.e., feature extraction, converting from RGB to a computational vector).  
The Transformer encoder then applies self-attention over all patch tokens, meaning each patch attends to every other patch to capture their relationships. 
This produces an attention-weighted set of patch features enriched with global context (visualized as a patch-to-patch attention map in Fig.~\ref{fig:background_detr}(b)). 
As a result, even distant parts of the image can influence each other’s features, helping the model recognize large-scale structures as well as contextual cues for object detection. 

After encoding, DETR’s Transformer decoder produces the object detections by interacting with a fixed set of learned object queries. 
These object queries can be thought of as placeholders for potential objects in the image. 
In the decoder, each query attends to the encoded patch features (through encoder–decoder cross-attention) and gradually updates its representation to focus on a specific object.
In Fig.~\ref{fig:background_detr}, the object queries (A, B, C, and D in Fig.~\ref{fig:background_detr}(c)) attend to different patches (illustrated by patch–query attention maps), and each query outputs a bounding box with a class label (e.g., car or pedestrian) and a confidence score. 
High-confidence detections (boxes of A, B, C  in Fig.~\ref{fig:background_detr}(d)) typically correspond to prominent objects, while low-confidence outputs (a box of D  in Fig.~\ref{fig:background_detr}(d)) may indicate a background or small objects that the model is less certain about. 

\subsection{Observation}
\label{subsec:observation}

To design a DETR that satisfies both timing guarantee (R1) and high accuracy (R2), we first examine how patch granularity affects DETR’s performance. 
We conducted a case study using the DINO~\cite{ZLL22}, a state-of-the-art DETR model, on the KITTI dataset~\cite{kitti12}, evaluating performance on an \hb{NVIDIA Jetson Orin} system-on-chip (SoC)~\cite{Orin}. 
From this study, two key empirical observations O1 and O2 emerge.

\begin{itemize}%[leftmargin=*]
    \item[O1.] Latency–accuracy trade-off: There is a clear trade-off between detection accuracy and inference speed depending on the patch size.
\end{itemize}

\begin{figure}[t] 
    \centering
    \subfloat[Latency–accuracy trade-off]{\includegraphics[width=0.48\linewidth]{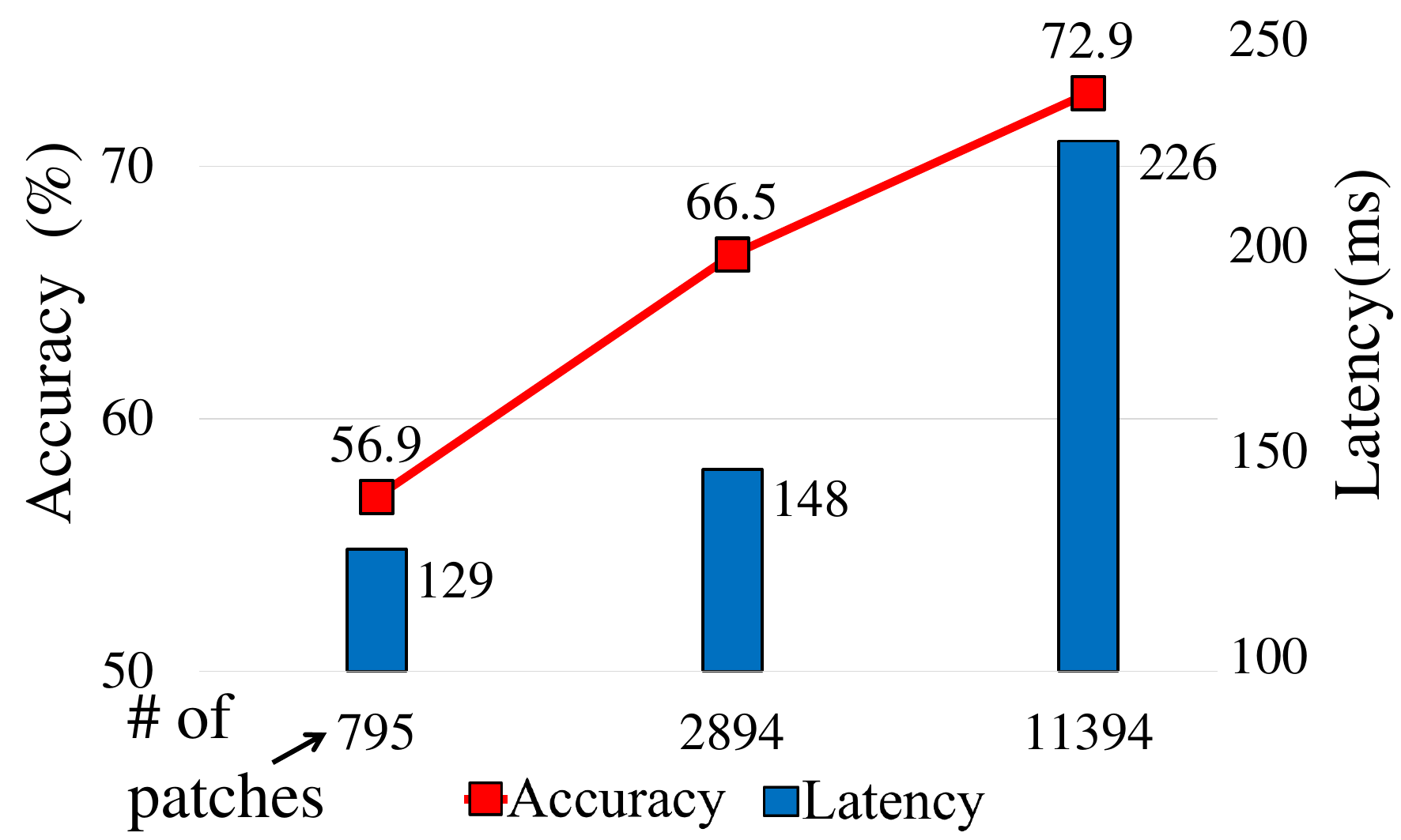}} 
    \subfloat[Accuracy by object size]{\includegraphics[width=0.48\linewidth]{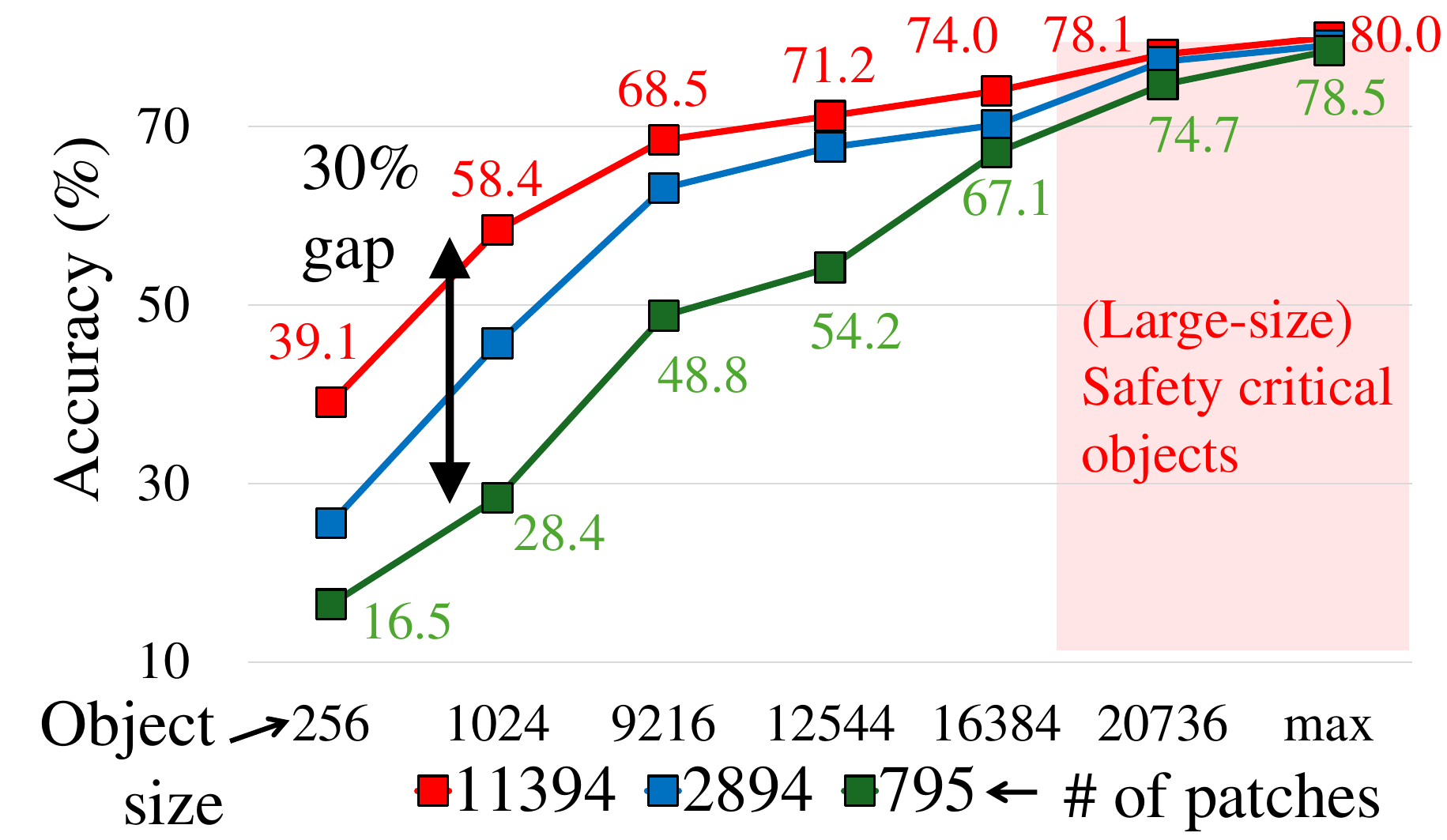}} \\
    \caption{Accuracy and latency analysis of DETR (DINO model) with varying patch granularity on the KITTI dataset (evaluated on Jetson Orin). (a) Trade-off between accuracy (averaged across all images) and latency with different numbers of patches. (b) Accuracy (averaged for each object-size group) comparison by object size, showing coarse patches are sufficient for large objects but significantly reduce accuracy for smaller objects.}
    \label{fig:motivation}    
\end{figure}

Fig.~\ref{fig:motivation}(a) illustrates that using coarser patches (e.g., the number of 795) yields faster inference (e.g., 129 ms) but at the cost of lower detection accuracy (averaged across all input images). 
Using finer patches (e.g., the number of 11394) improves accuracy, but incurs much higher latency (e.g., 226 ms). 
This indicates that %For example, 
an extremely coarse division (very few patches) might run quickly but only detect large, obvious objects (compromising R2), whereas a very fine division greatly boosts accuracy on small objects but could violate the real-time latency requirements (compromising R1). 
This fundamental trade-off implies that no single patch resolution can simultaneously satisfy both R1 and R2.

\begin{itemize}%[leftmargin=*]
    \item[O2.] Object size vs. patch granularity: Large, safety-critical objects are reliably detected even with coarse patches, whereas small objects require finer granularity for high accuracy.
\end{itemize}

Fig.~\ref{fig:motivation}(b) supports this insight by breaking down accuracy according to object size, with the x-axis representing object area (width$\times$height pixels). 
\hb{DETR} using coarse patches achieves accuracy (averaged per object-size group) nearly comparable to that with fine patches for large objects, such as nearby vehicles occupying an area exceeding 16384 pixels. 
However, accuracy significantly deteriorates for smaller objects, showing roughly a 30\% accuracy gap between coarse (green curve) and fine settings (red curve). This confirms that coarse patches are sufficient for large objects, but finer patches are required to reliably detect smaller ones.

These observations highlight a fundamental challenge O1 in employing DETR for AV perception, while also providing a critical insight O2 toward overcoming it. Specifically, high accuracy is achievable with coarse patches (at low latency) for images containing only large objects. For images featuring varying object sizes, selectively applying coarse patches to large-object regions and fine patches to small-object regions maintains high accuracy while incurring lower latency compared to uniformly using fine patches.

%CF-DETR 
\sys{} systematically exploits this DETR-specific property and employs specialized batching techniques A1--A3 to navigate the accuracy-latency trade-off. Furthermore, it integrates a scheduling framework A4 ensuring timing guarantees. 
These system-level design elements are detailed in the following section.

%\section{System Design} %\hbcmt{well-known하거나 우리의 main contribution이 아닌 부분은 간략히 설명할 것.}

\begin{figure*}[t]  % *를 붙이면 두 개의 column을 다 사용
    \centering
    \includegraphics[width=0.85\linewidth]{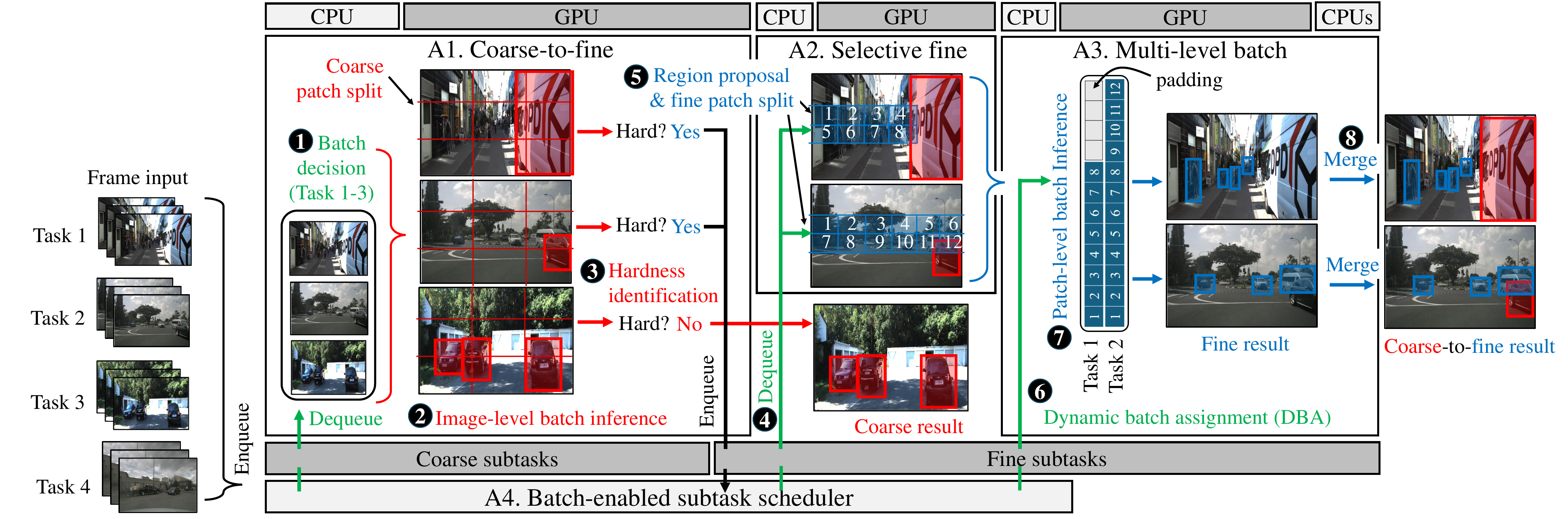}
    \caption{System overview of \textsf{CF-DETR}: Periodic frames from multiple tasks arrive and queue. The scheduler then forms initial processing batches (\ding{182}).
    Inputs undergo coarse batch inference using a low-res grid—yielding initial detected boxes through concurrent DETR inferences in a single batch (\ding{183}).
    Frames are classified by confidence as \textit{easy} (high-confidence) or \textit{hard} (ambiguous) (\ding{184}); this determines if fine subtasks are needed. 
    Easy frames use coarse results directly.
    Hard frames generate fine subtasks, which are queued for refinement. The scheduler determines their execution timing (\ding{185}).
    For each fine subtask, hard regions are identified and subdivided into finer patches, defining its workload (\ding{186}).
    The scheduler selects an dynamic fine batch configuration from active subtasks, maximizing throughput within deadlines (\ding{187}).
    Fine subtasks use the patch-level batch inference to refine detections in uncertain regions (\ding{188}).
    Finally, refined outputs merge with high-confidence coarse detections for the final results (\ding{189}). 
    }
    \label{fig:system_overview}
\end{figure*}

\section{SYSTEM DESIGN}
\label{sec:system_design}

This section details the design of \sys{}, encompassing its core architectural strategies and the real-time scheduling framework, followed by experimental validation of its key components' functionality.
%and overall effectiveness in meeting stringent timing and accuracy demands for multi-DETR execution.

\subsection{Overview}
\label{subsec:system_overview}

To satisfy both real-time and accuracy requirements R1 and R2, \sys{} leverages insights from our experimental observations in Sec~\ref{subsec:observation} and employs four key synergistic strategies: (A1) coarse-to-fine inference, (A2) selective fine inference, (A3) multi-level batch inference, and (A4) a batch-enabled subtask scheduler. 
%These strategies are designed to dynamically manage computational resources, effectively balancing detection performance with latency. 
The overall system architecture and workflow, which demonstrates how these strategies process incoming frames from multiple tasks---initiating with a coarse subtask, followed by adaptive fine-grained processing and batching, all orchestrated by the real-time scheduler---are depicted in Fig.~\ref{fig:system_overview}. 
The detailed design of these components is elaborated in Sec.~\ref{subsec:system_design}.

\textbf{Benefits.}
Leveraging its coarse-to-fine paradigm and integrated scheduling, \sys{} offers several key benefits. 
First, its broad applicability allows compatibility with most (if not all) Transformer-based object detectors by utilizing common architectural components. 
Second, it ensures prioritized responsiveness and accuracy for large, safety-critical objects, which are typically detected rapidly and reliably in the coarse stage (A1), crucial for safety in autonomous systems. 
Third, \sys{} maximizes overall detection accuracy while adhering to deadlines by adaptively scaling fine-stage computation (A2) based on available slack time. 
Finally, it enhances GPU throughput via multi-level batch inference (A3), a specialized batch processing scheme for Transformer computations, effectively parallelizing inference across multiple tasks.

\subsection{Design of \sys{}}
\label{subsec:system_design}
 
We now detail the design and empirical validation of the key components of \sys{}. The system dedicates a thread for each camera and another for the scheduler, with inter-thread communication facilitated by shared memory. Scheduler operations are performed on the CPU, while DNN inference is executed on the GPU.

\textbf{A1. Coarse-to-fine inference (frame-level: \ding{183}--\ding{184}).}
Incoming frames periodically generate coarse subtasks, which are enqueued. The scheduler determines their execution timing and whether to process them individually or in a image-level batch. These subtasks perform rapid detection using coarse-grained patches (with a predefined granularity) (\ding{183}). This initial inference yields, for each object query, a candidate object's location ($x,y$), size ($w,h$), and confidence score ($c$, where $0 \le c \le 1$).
Based on these results, frames are classified by their difficulty (\ding{184}). Specifically, after excluding objects with high confidence (e.g., $c > 0.8$), typically deemed large and safety-critical, an average confidence score is computed for the remaining candidates within each frame. If this average is below a certain threshold (e.g., 0.05), the frame is classified as ``easy''; its coarse results are considered sufficient, and fine-grained processing is skipped. Such a low average implies that, after high-confidence critical objects are accounted for, the remaining detections predominantly correspond to background regions. Conversely, frames with higher average confidence scores (indicating more ambiguity or complexity, often due to smaller or less distinct non-critical objects) are labeled ``hard,'' triggering selective refinement. 
%Coarse-stage feature tokens are cached for reuse.

\begin{figure*}[t] % Use regular 'figure' for one-column width
    \centering
    % First row of subfigures
    \subfloat[Query results of coarse inference.]{\includegraphics[width=0.24\linewidth]{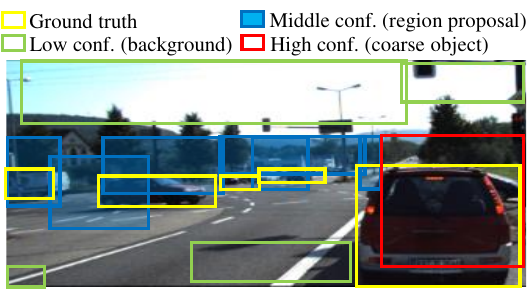}} 
    \vspace{0.1cm}
    \subfloat[A1: Query confidence for easy/hard frames.]{\includegraphics[width=0.24\linewidth]{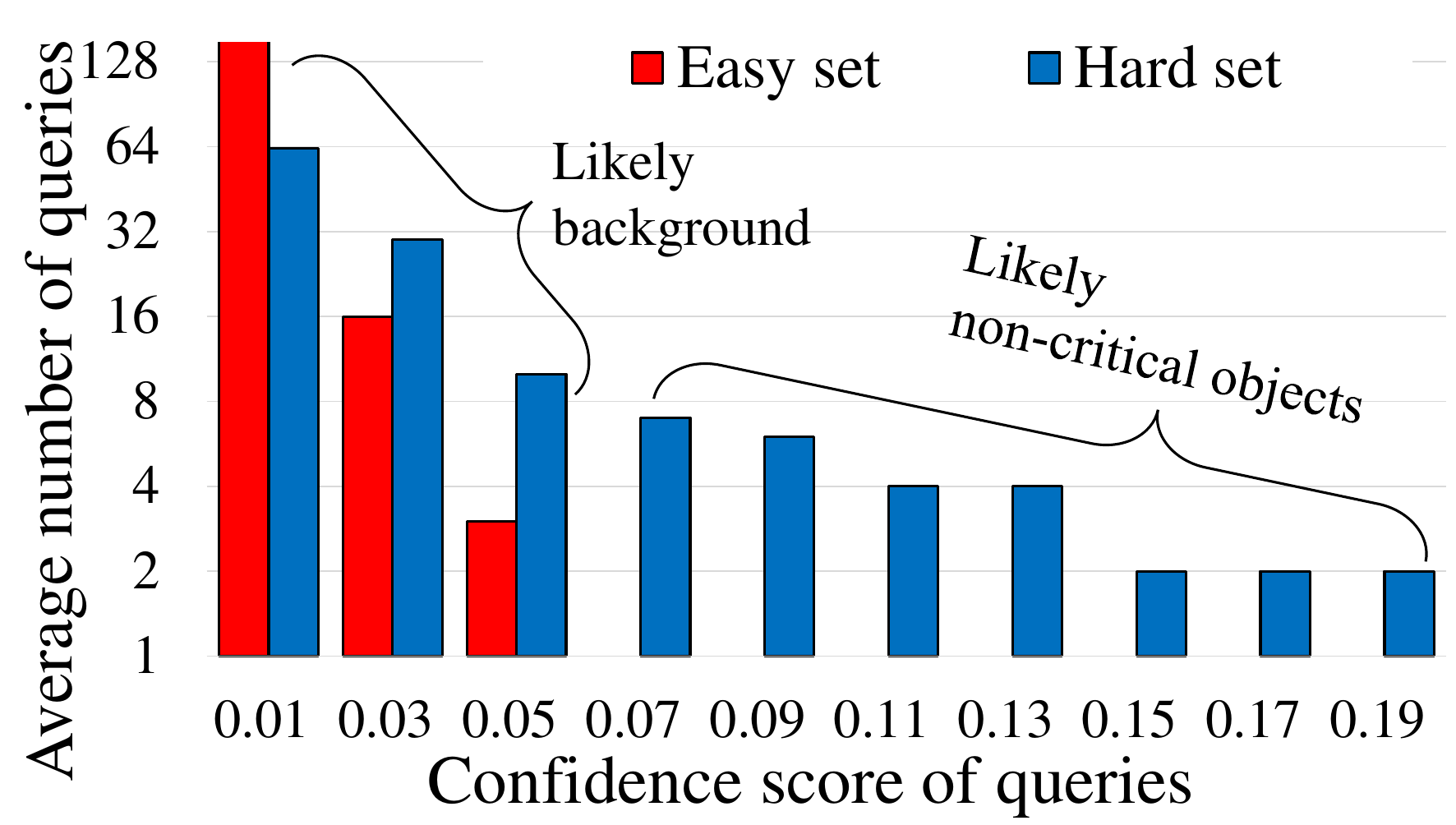}}    
    \vspace{0.1cm}
    \subfloat[A2: Fine-resolution paches and coverage.]{\includegraphics[width=0.24\linewidth]{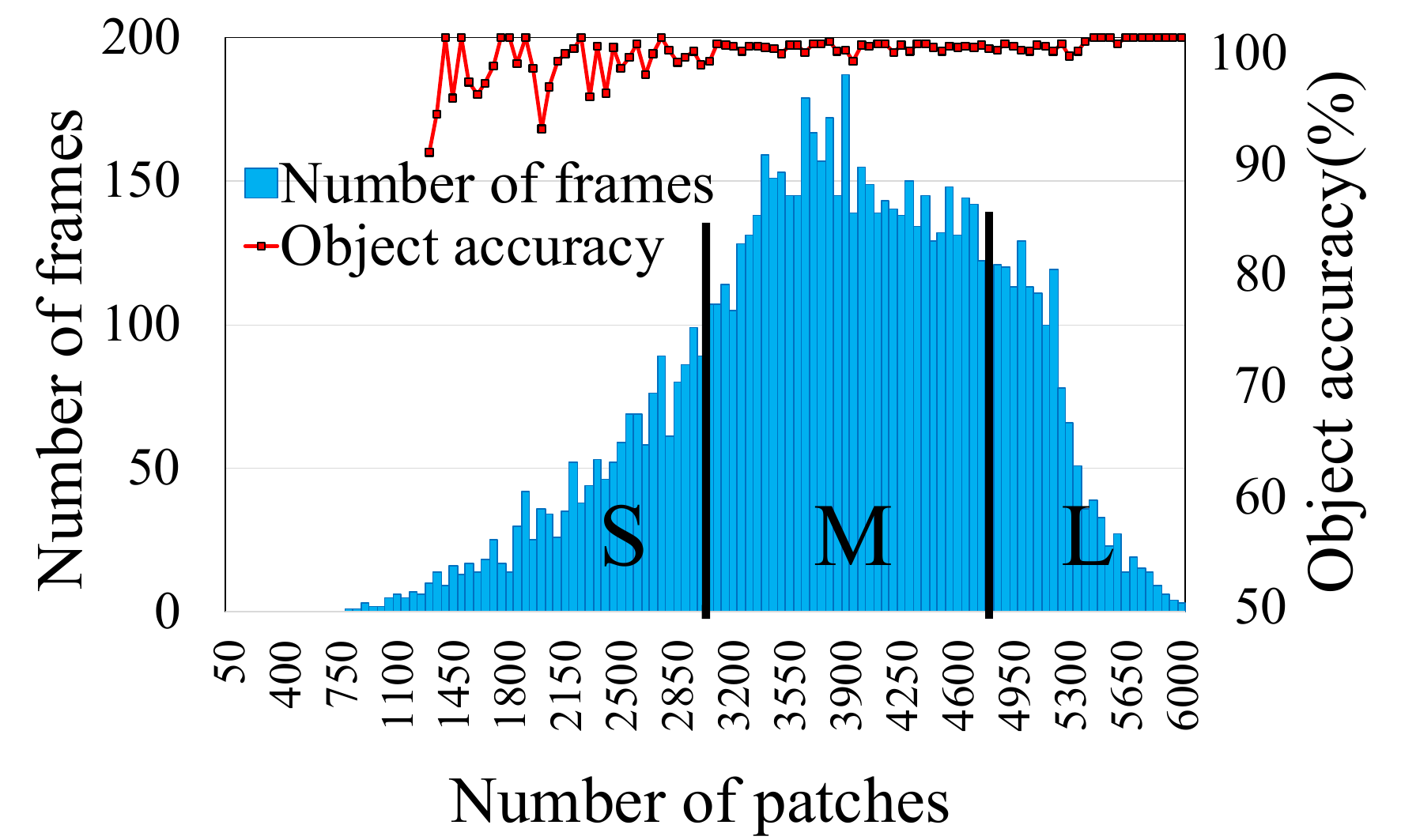}}
    \vspace{0.1cm}
    \subfloat[A3: Fine-stage latency with multi-level batch inference.]{\includegraphics[width=0.24\linewidth]{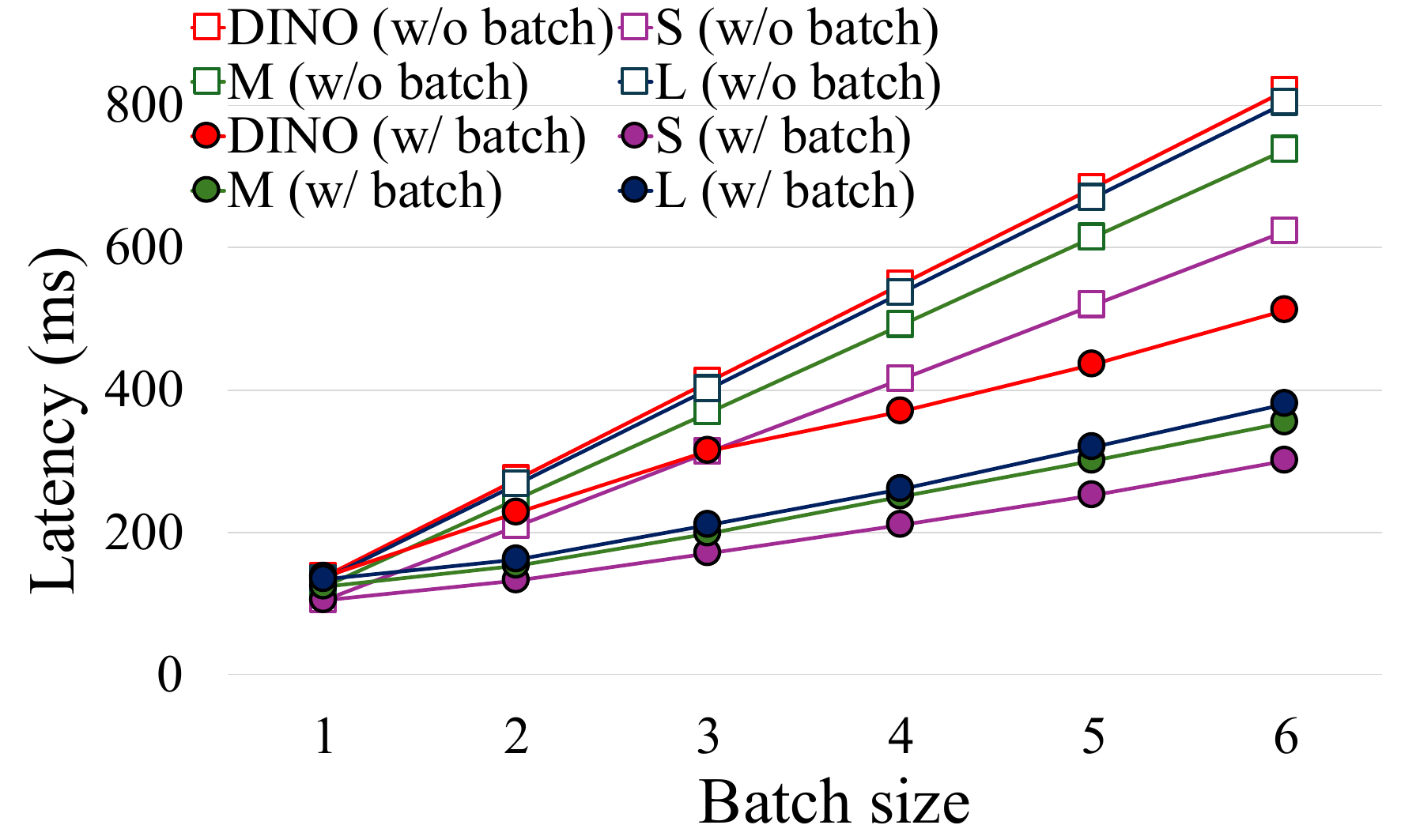}}    
    \caption{\hb{Validation of \sys{}'s adaptive inference strategies A1--A3 using the DINO model on Jetson Orin with the KITTI dataset: (a) sample coarse inference output; (b) hardness identification by query confidence analysis; (c) selective fine inference efficiency, showing patch count versus object coverage; and (d) multi-level batch inference latency advantages for fine-stage processing.}}
\label{fig:validation}
\end{figure*}

\textbf{Validating \ding{184}.}
Fig.~\ref{fig:validation}(a) visualizes a sample coarse inference output (DINO model, 795 patches, KITTI), with color-coded boxes indicating ground truth (yellow), high-confidence queries ($c>0.8$, red) for prominent objects, intermediate-confidence queries ($0.05 < c \le 0.8$, blue) flagged for refinement, and background-indicative queries ($c \le 0.05$, green). 
To determine frame difficulty, we analyzed object query confidence distributions across KITTI after coarse inference and filtering of high-confidence queries, defining an ``easy set" (frames with $\le 5$ remaining intermediate-confidence queries) and a ``hard set" (frames with $>10$). 
As shown in Fig.~\ref{fig:validation}(b), the average query confidence distributions distinctly separate easy set (highly skewed, near-zero confidences) from hard set (flatter distributions with more intermediate confidences); this distinction enables \sys{} to trigger fine inference selectively and underpins our use of an average confidence threshold (e.g., 0.05) to classify frames as easy or hard.

\textbf{A2. Selective fine inference (region-level: \ding{186}).}
For frames classified as ``hard'' by A1, \sys{} performs region proposal and fine patch split (\ding{186}), concentrating computation on challenging regions of interest (ROIs) instead of the entire image. 
These ROIs are proposed based on the locations ($x,y$) and sizes ($w,h$) of intermediate-confidence queries from the A1 coarse inference stage. 
Critically, only these query-defined regions are subsequently re-partitioned into finer patches, which are then used to process the selected ROIs, thereby enhancing local resolution. 
Meanwhile, non-ROI areas reuse the original coarse-stage tokens from A1. 

\textbf{Validating \ding{186}.}
Fig.~\ref{fig:validation}(c) illustrates the distribution of fine-resolution patch counts generated by \sys{}'s region proposal and selective patch split mechanism (\ding{186}). 
This performance is benchmarked against a DINO model that typically requires more than 10,000 patches for equivalent full-frame fine resolution. 
The figure also plots corresponding object accuracy (\%), indicating the ground-truth coverage within these selectively refined regions. 
Consequently, \sys{} processes \jl{the hardest} %most hard 
frames with under 6,000 fine-resolution patches—a mere fraction of DINO's more than 10,000 for equivalent full resolution—achieving a significant latency advantage yet maintaining comparable accuracy by capturing nearly 100\% of ground-truth objects in these selectively refined areas.

\textbf{A3. \textit{Multi-level batch inference} (batch-level: \ding{188}--\ding{189}).}
To enable latency-effective batched execution of fine-grained patches derived from A2-identified hard frames, \sys{} introduces \textit{patch-level} batch inference. 
This approach avoids the accuracy degradation common in traditional \textit{image-level} batching (arising from input cropping/resizing to meet identical input size constraints, which can cause information loss) by leveraging A2's already feature-extracted fine patches. 
These patches originate from A2's validated region proposals (\ding{186}), which themselves exhibit high ground-truth coverage (per Fig.~\ref{fig:validation}(c)). 
Since patch counts from A2 can vary across tasks, padding is applied to equalize them for batch configuration and assignment (\ding{188}) before they are processed in the subsequent patch-level batch inference step (\ding{189}). 
This patch-level method offers two key advantages over image-level batching: 1) it eliminates redundant backbone feature re-extraction (cf. Fig.~\ref{fig:background_detr}(a)), and 2) it achieves substantial latency improvements by processing only these targeted, high-coverage patches.

\textbf{Validating \ding{188}.}
Patch-level batch inference yields substantial latency benefits, as shown in Fig.~\ref{fig:validation}(d). This figure illustrates that compared to processing fine tasks individually (empty squares), batched execution (circles) significantly reduces total latency; for instance, six concurrent fine inferences achieve an approximate $2\times$ speedup.
Furthermore, Fig.~\ref{fig:validation}(d) compares the inference times of conventional image-level batching for DINO (red circles) against \sys{}'s patch-level batch inference (after region proposal \ding{186}). 
For this comparison, patches derived from hard frames \jl{were evaluated, based on analysis related to Fig.~\ref{fig:validation}(c), where patch counts like 3000 and 4800 were used as thresholds to categorize into small (S), medium (M), and large (L) sets for simplicity.}
%For this comparison, patches derived from hard frames (based on analysis related to Fig.~\ref{fig:validation}(c), where patch counts like 3000 and 4800 were used as thresholds to categorize into low (L), medium (M), and high (H) sets for simplicity) were evaluated. 
The S, M, and L sets (represented by purple, green, and blue circles, respectively, in Fig.~\ref{fig:validation}(d)) show significant improvements: at a batch size of six, DINO's image-level approach took approximately 500\,ms, whereas \sys{}'s patch-level processing required only around 300\,ms for S-set patches and 400\,ms for L-set patches, demonstrating latency reductions of approximately 20--40\%.

\textbf{A4. Batch-enabled subtask scheduler (\ding{182}, \ding{185}, and \ding{187}).}
%A real-time scheduler orchestrates the coarse and fine processing stages across multiple camera tasks, employing a non-preemptive, fixed-priority policy. 
%Top priority is assigned to coarse inference, guaranteeing initial detection within its deadline. 
%Fine tasks are then scheduled in the remaining slack time, enhancing accuracy without jeopardizing coarse task timeliness. 
The scheduler is batch-aware: it forms initial coarse processing batches (\ding{182}) and also handles the dynamic assignment of fine tasks into batches (\ding{187}) for execution via A3 when feasible. 
Operationally, incoming frames are dispatched for coarse processing. 
Hard frames identified by A1 generate fine subtasks, which are queued. 
When the GPU is free from high-priority coarse work, the scheduler determines their execution timing (\ding{185}) and executes these pending fine subtasks, often as optimally configured batches (utilizing \ding{187}) from the queue. 
This coordination ensures critical deadlines are met while maximizing timely fine-grained refinements. Sec.~\ref{sec:scheduling_framework} details this scheduler's operation.

\section{Scheduling Framework}\label{sec:scheduling_framework}

%\jlcmt{strictly periodic임을 어딘가에 서술}

This section presents the task model and the \npfpstar{} scheduling framework developed for \sys{}. 

\subsection{Task model}\label{subsec:task_model}
We consider a multi-object detection system comprising $n$ strictly-periodic DETR tasks, $\tau = \{\tau_1, \ldots, \tau_n\}$. 
These tasks execute on a heterogeneous platform (multiple CPUs, single shared GPU), such that their CPU and GPU processing components execute exclusively. 
\hb{Where advantageous, parallelizable CPU operations can utilize multi-core threading, while accounting for inter-core communication costs (e.g., \ding{189} in Fig.~\ref{fig:system_overview}).}
Each task $\tau_i \in \tau$ is defined by its period $T_i$ and a relative deadline $D_i$; we assume $D_i = T_i$ for all tasks.

A central concept of the \sys{} model is the partitioning of each task $\tau_i$ into two sequential components: a safety-critical coarse subtask $\tau_i^S$, followed by an optional fine subtask $\tau_i^F$. 
Formally, the task structure is represented as $\tau_i = (\tau_i^S, \tau_i^F, T_i, D_i)$.
The subtasks are parameterized as follows:
\begin{itemize}[leftmargin=*]
    \item The coarse subtask $\tau_i^S = (p^S, C_i^S)$ processes a fixed, pre-determined number of coarse patches, $p^S$, with a worst-case execution time (WCET) of $C_i^S$. The decision to subsequently invoke the fine subtask $\tau_i^F$ is made during the execution of $\tau_i^S$, specifically within its hardness determination process (cf. \ding{184} in Fig.~\ref{fig:system_overview}).
    \item The fine subtask $\tau_i^F = (p_i^F, C_i^F(p_i^F))$ processes a dynamically determined number of fine patches, $p_i^F$, which can be one of several predefined levels (e.g., S, M, or L, corresponding to $0 < \text{S} < \text{M} < \text{L}$ patches), or zero if the fine subtask is skipped. The value of $p_i^F$ is identified during the selective patch splitting stage (cf. \ding{186} in Fig.~\ref{fig:system_overview}) performed at the beginning of $\tau_i^F$ itself. Consequently, its WCET, $C_i^F(p_i^F)$, is contingent upon this chosen $p_i^F$. If no fine processing is required, $\tau_i^F$ is skipped (effectively $p_i^F=0$, and $C_i^F(0)=0$).
\end{itemize}
The WCETs, $C_i^S$ and $C_i^F(p_i^F)$, are calculated based on their constituent operations:
\begin{align}
    C_i^S &= c_i^{ps} + c_i^{at}(p^S) + c_i^{dt}, \label{eq:wcet_coarse} \\
    C_i^F(p_i^F) &= c_i^{sps}(p_i^F) + c_i^{at}(p_i^F), \label{eq:wcet_fine}
\end{align}
where $c_i^{ps}$ is the WCET for coarse patch splitting, $c_i^{at}(p)$ is for the attention mechanism over $p$ patches ($p=p^S$ for $\tau_i^S$, $p=p_i^F$ for $\tau_i^F$), $c_i^{dt}$ is for hardness determination, and $c_i^{sps}(p_i^F)$ is for selective patch splitting (which itself depends on the outcome that determines $p_i^F$).
Task periods $T_i$ may differ (e.g., for different camera types), though the model also applies to uniform periods. A job is considered active at time $t$ if it has been released but not yet completed. We use $\tau^S(t)$ and $\tau^F(t)$ to denote the sets of active coarse and fine subtasks at time $t$, respectively, and $r_i(t)$ to denote the next release time of task $\tau_i$ after or at $t$. For periodic tasks, release times and deadlines are deterministic. 
The coarse subtask $\tau_i^S$ must be completed by its deadline $D_i$; if its corresponding fine subtask $\tau_i^F$ is subsequently initiated, it executes after $\tau_i^S$ and must also conclude by this same deadline $D_i$.

The WCET values for individual coarse subtasks ($C_i^S$) and fine subtasks ($C_i^F(p_i^F)$), along with those for their batched executions ($C_{\mathcal{B}^S}$ and $C_{\mathcal{B}^F}$, where $\mathcal{B}^S$ and $\mathcal{B}^F$ are sets of batched coarse and fine subtasks, respectively, with $|\mathcal{B}^S|\ge2$ and $|\mathcal{B}^F|\ge2$), are determined through offline, measurement-based WCET analysis, as detailed in Sec.~\ref{sec:evalution}. 
As a property of batch execution, we assume $C_{\mathcal{B}^S} \le \sum_{\tau_i\in\mathcal{B}^S} C_i^S$ (and similarly for $C_{\mathcal{B}^F}$). 
\sys{} exclusively relies on these pre-determined offline WCETs, not on runtime predictions. 
Achieving predictable WCETs is feasible given uniform input sizes and dedicated GPU utilization per DNN type, which avoids resource contention~\cite{KCK22, XYK19, KLC22}. 
This methodology aligns with recent advancements in WCET analysis for DNNs~\cite{KLH24}. We operate under the assumption that a reliable upper bound on WCET can be established through intensive measurements complemented by appropriate safety margins.

\setlength{\textfloatsep}{10pt}
\begin{algorithm} [t]
\caption{The NPFP$^{**}$ generic algorithm} 
\label{algo:npfp_star}
\small
\raggedright
At scheduling instant $t$ (job completion or arrival on idle system): 
\begin{algorithmic}[1] % Added line numbers
    \IF {$\tau^S(t) \neq \emptyset$} % Check if multiple coarse subtasks are active 
        \IF {The first ``*'' in NPFP** represents \textsf{\footnotesize C}}
        %\IF {\textsf{C} $\in$ **}
            \STATE Execute the highest-priority active coarse task in $\tau^S(t)$.
        \ELSIF {The first ``*'' in NPFP** represents \textsf{\footnotesize [C]}}
        %\ELSIF {\textsf{[C]} $\in$ **}
            \STATE Execute a coarse batch using Algo.~\ref{algo:coarse_batch}.
            %\jlcmt{(possibly a single highest-priority task)써야 할까}
        \ENDIF
    \ELSIF{$\tau^F(t) \neq \emptyset$}
        %\STATE \jlcmt{시간이 모자라면 F 실행을 안함이 서술 되어 있는가?}
        \IF {The second ``*'' in NPFP** represents \textsf{\footnotesize F}}
        %\IF {\textsf{F} $\in$ **}
            \STATE Execute the highest-priority active fine task in $\tau^F(t)$ if its WCET plus current time $t$ is no later than the earliest future release of any coarse subtask.
        \ELSIF {The second ``*'' in NPFP** represents \textsf{\footnotesize [F]}}
        %\ELSIF {\textsf{[F]} $\in$ **}
            \STATE Execute fine batch(s) using Algo.~\ref{algo:fine_batch}.
        \ENDIF
    \ENDIF
\end{algorithmic}
\normalsize
\end{algorithm}

\subsection{The \npfpstar{} framework}
\label{subsec:npfp_star_framework}

%\jlcmt{Algo 1에서 include라 표현하면 C는 항상 [C]에 포함인데... ``the first * is C'' 이렇게 해야 하나?}

The \npfpstar{} framework establishes a unified non-preemptive fixed-priority (NPFP) scheduling approach for the \sys{} system. To delineate the various execution strategies within this framework, we adopt the following notation for the superscript of NPFP: \textsf{\small C} and \textsf{\small F} denote the individual (non-batched) execution of coarse ($\tau_i^S$) and fine ($\tau_i^F$) subtasks, respectively. When enclosed in brackets, \textsf{\small [C]} and \textsf{\small [F]} signify the batched execution of coarse and fine subtasks, respectively. 
This framework accommodates these different modes while aiming to ensure timing guarantees. Given that fine subtask execution (\textsf{\small F} or \textsf{\small [F]}) is optional and opportunistic, several instantiations of the \npfpstar{} framework can be realized, such as \mbox{NPFP$^\textsf{C}$} (\textsf{\small C}-only), \mbox{NPFP$^\textsf{CF}$}, \mbox{NPFP$^\textsf{[C]F}$}, \mbox{NPFP$^\textsf{C[F]}$}, and \mbox{NPFP$^\textsf{[C][F]}$}.
While individual subtasks (either \textsf{\small C} or \textsf{\small F}) run non-preemptively once initiated, preemption by a higher-priority \textit{task} is permitted, but only at the distinct boundary between the completion of a task's coarse subtask and the potential initiation of its fine subtask.

\textbf{Principle.}
The \npfpstar{} framework is guided by \sys{}'s core execution principles: coarse subtasks address safety-critical detections, while fine subtasks opportunistically enhance overall accuracy. 
A fundamental rule is that coarse subtasks ($\tau_i^S$) always possess higher scheduling priority than fine subtasks ($\tau_i^F$). 
This ensures that safety-critical processing is prioritized; consequently, fine subtasks can only commence execution when no coarse subtasks are ready. 
Among coarse subtasks (likewise fine subtasks), a pre-defined priority is assigned.
The offline schedulability guarantee primarily covers the individual execution of all coarse subtasks (\textsf{\small C}). 
Decisions regarding batched coarse execution (\textsf{\small [C]}), or any fine subtask execution (either individual, \textsf{\small F}, or batched, \textsf{\small [F]}), are made dynamically at runtime. These runtime choices aim to maximize accuracy by utilizing available resources, always contingent upon not compromising the pre-established schedulability of the critical coarse subtasks; critically, the overhead of any runtime schedulability analysis required for these decisions must be \textit{minimal}.
%\jlcmt{run-time overhead가 (1),(2)번식에 반영되나?}

Algo.~\ref{algo:npfp_star} illustrates the generic decision logic within the \npfpstar{} framework at each scheduling instant $t$. 
It strictly prioritizes active coarse subtasks ($\tau^S(t)$) over active fine subtasks ($\tau^F(t)$). 
If any coarse subtasks are active (Line~1), a decision is made according to the specific NPFP policy variant (denoted by ``**''). Depending on the policy, the system either selects the highest-priority individual coarse subtask (Line~3, if the policy includes \textsf{\small C}) or invokes Algo.~\ref{algo:coarse_batch} to schedule a batch of coarse subtasks (Line~5, if the policy includes \textsf{\small [C]}). Note that Algo.~\ref{algo:coarse_batch} may still choose to execute only the highest-priority active coarse subtask, depending on its run-time schedulability test result.
%If coarse subtasks are active (Line~1), a decision is made, based on the specific \npfpstar{} policy variant (indicated by ``**''), to execute either the highest-priority individual coarse subtask (Line~3, if the policy includes \textsf{\small C}) or a batch of coarse subtasks \jl{by Algo.~\ref{algo:coarse_batch} to be presented} (Line~5, if the policy includes \textsf{\small [C]}); \jl{note that Algo.~\ref{algo:coarse_batch} may decide to execute the higher-priority active coarse task only due to its run-time schedulability test.}
%(Line~5, if the policy includes \textsf{\small [C]}, utilizing a defined batching algorithm such as Algo.~\ref{algo:coarse_batch} for coarse batching to be presented). 
If no coarse subtasks are active but fine subtasks are (Line~7), a similar policy-dependent decision is applied: either executing the highest-priority individual fine subtask (Line~9, if the policy includes \textsf{\small F}, contingent on its WCET being less than or equal to the time until the earliest future coarse subtask release) or a batch of fine subtasks (Line~11, if the policy includes \textsf{\small [F]}, employing a dynamic fine batch assignment strategy in Algo.~\ref{algo:fine_batch} to be executed). 
The specific conditions guiding the choice between individual and batched execution (represented by ``**'') and the referenced batch formation algorithms are defined by the particular \npfpstar{} instantiation (e.g., \npfpcbf{} or \npfpcbfb{}). 
The foundational analysis of \npfpc{}, the simplest variant involving only individual coarse subtask execution, establishes the baseline for system schedulability guarantees.

Following NPFP rules, scheduling decisions occur at discrete instants $t$, triggered by job completions (if other jobs are pending) or new job arrivals on an idle platform. 
At each such point $t$, the highest-priority ready coarse subtask $\tau_i^S$ from the set of active coarse subtasks $\tau^S(t)$ is selected and runs non-preemptively for its duration $C_i^S$. 
The offline schedulability for \npfpc{} is given by Lemma~\ref{lemma:offline}.

\begin{lemma}[NPFP schedulability test~\cite{BTW95,YBB10,ChBr17a}] \label{lemma:offline}
A task set $\tau$ scheduled by \npfpc{} is schedulable, if every $\tau_i \in \tau$ satisfies Eq.~\eqref{eq:rta1}.
%%$R_i \le T_i$, 
%where $R_i$ is its worst-case response time determined as follows.

\begin{align}\label{eq:rta1}
R_i \le T_i,
\end{align}
where the worst-case response time $R_i$ is the smallest positive value found by iterating Eq.~\eqref{eq:rta2} until convergence,
i.e., $R_i(x+1)=R_i(x)$, starting with $R_i(0) = C_i^S + B_i$:
\begin{align}\label{eq:rta2}
R_i(x+1) = C_i^S + B_i + \sum_{\tau_h \in \texttt{HP}(\tau_i)} \bigg\lceil \frac{R_i(x)}{T_h} \bigg\rceil \cdot C_h^S.
\end{align}
Here, $B_i = \max ( \{0\} \cup \{ C_j^S \mid \tau_j \in \texttt{LP}(\tau_i) \} )$ denotes the maximum blocking from lower-priority tasks of $\tau_i$ ($\texttt{LP}(\tau_i)$), and $\texttt{HP}(\tau_i)$ represents the set of higher-priority tasks of $\tau_i$.
\end{lemma}
\begin{IEEEproof}
We analyze a level-$i$ busy period. Without loss of generality, let this period start at $t=0$ with job $J_i$ of task $\tau_i$. 
The critical instant theorem~\cite{GRS96} (Lemma 6) states $\tau_i$'s worst-case response time occurs with synchronous higher-priority ($\texttt{HP}(\tau_i)$) arrivals at $t=0$, after maximum blocking $B_i = \max ( \{0\} \cup \{ C_j^S \mid \tau_j \in \texttt{LP}(\tau_i) \} )$ from a lower-priority task in $\texttt{LP}(\tau_i)$.
In an interval $[0, L)$, the higher-priority task workload is at most $\sum_{\tau_h\in\texttt{HP}(\tau_i)} \lceil\frac{L}{T_h}\rceil\cdot C_h^S$. Eq.~\eqref{eq:rta2}'s fixed point, $R_i$, sums $C_i^S$, $B_i$, and higher-priority interference, bounding the time to finish $J_i$. Thus, if $R_i \le T_i$ (Eq.~\eqref{eq:rta1}), $J_i$ meets its deadline. This establishes the base case.

For the inductive step, we show job separation $t_{x+1} - t_x \le T_i$ within the busy period. At $t_x$ (the start time of the $x^{th}$ job), no higher-priority jobs are pending. The time for $C_i^S$ plus higher-priority interference within $[t_x, t_x+L)$ is $R_i'(L) = C_i^S+\sum_{\tau_h\in\texttt{HP}(\tau_i)} \lceil\frac{L}{T_h}\rceil\cdot C_h^S$, which omits $B_i$. Since $R_i'(L)$ is bounded by the value from Eq.~\eqref{eq:rta2} (including $B_i$), the premise $R_i \le T_i$ ensures the solution $L^*$ to $L=R_i'(L)$ also satisfies $L^* \le T_i$. This means the $x^{th}$ job's load resolves within $T_i$, allowing the $(x+1)^{th}$ job to start by $t_x + T_i$. Hence $t_{x+1} - t_x \le T_i$, consistent with anti-self-pushing results~\cite{YBB10}. 
The base and inductive steps prove the lemma by induction. 
\end{IEEEproof}
\vspace{5pt}

\subsection{Coarse batch assignment}
\label{subsec:coarse_batch}

%\hbcmt{Following job analysis는 안해도 되나?}\jlcmt{Lemma 2에서 대략 해결함.}

The coarse batch assignment mechanism is invoked when an \npfpstar{} policy supporting batched coarse execution (\textsf{\small [C]}) is active (e.g., as called by Line~5 of Algo.~\ref{algo:npfp_star}, referring to Algo.~\ref{algo:coarse_batch} detailed herein). 
Coarse batching may impact the schedulability guaranteed by Lemma~\ref{lemma:offline} for individual coarse subtasks. 
As runtime verification of schedulability can incur significant overhead, a core objective for any \npfpstar{} variant implementing \textsf{\small [C]} is to minimize this runtime analysis cost. 
Our approach achieves this by adopting a carefully designed batch policy with systematic constraints. 
While drawing inspiration from methods like Batch-MOT~\cite{KLH24}, we incorporate additional constraints that dramatically simplify the runtime analysis by ensuring batch execution aims for theoretical simplicity and low runtime cost. 
As demonstrated in Sec.~\ref{sec:evalution}, this constrained method achieves near-maximal accuracy, suggesting complex runtime batching analyses may primarily add overhead without proportional benefits.

\textbf{Principle.}
To minimize the runtime schedulability analysis cost, our design for coarse batching is founded on two key principles. 

\begin{itemize}
    \item [P1.] 
    %Only batches formed from a contiguous sequence of active coarse subtasks in $\tau^S(t)$, ordered from the highest to lowest priority, are considered permissible.
    Only batches consisting of a contiguous sequence of active coarse subtasks in $\tau^S(t)$,
    ordered from highest to lowest priority and always including the highest-priority task, are considered permissible.
    \item [P2.] Any coarse batch execution starting at time $t$ must complete before the earliest future release time of any task in the system, i.e., 

    \noindent
    \small
    \begin{align}\label{eq:coarse_batch_online} % This is Eq. 5 from the PDF (page 8)      
        C_{\mathcal{B}^S} \le \min_{\tau_i \in \tau^S} r_i(t) - t.
    \end{align}    
    \normalsize    
where $r_i(t)$ is the earliest release time of %task
$\tau_i$ after or at $t$. %\jlcmt{min에 $\tau^S(t)$에서 $(t)$ 빠져야 하지 않나?}
\end{itemize}

\begin{figure}[t]
    \centering
    \includegraphics[width=0.9\linewidth]{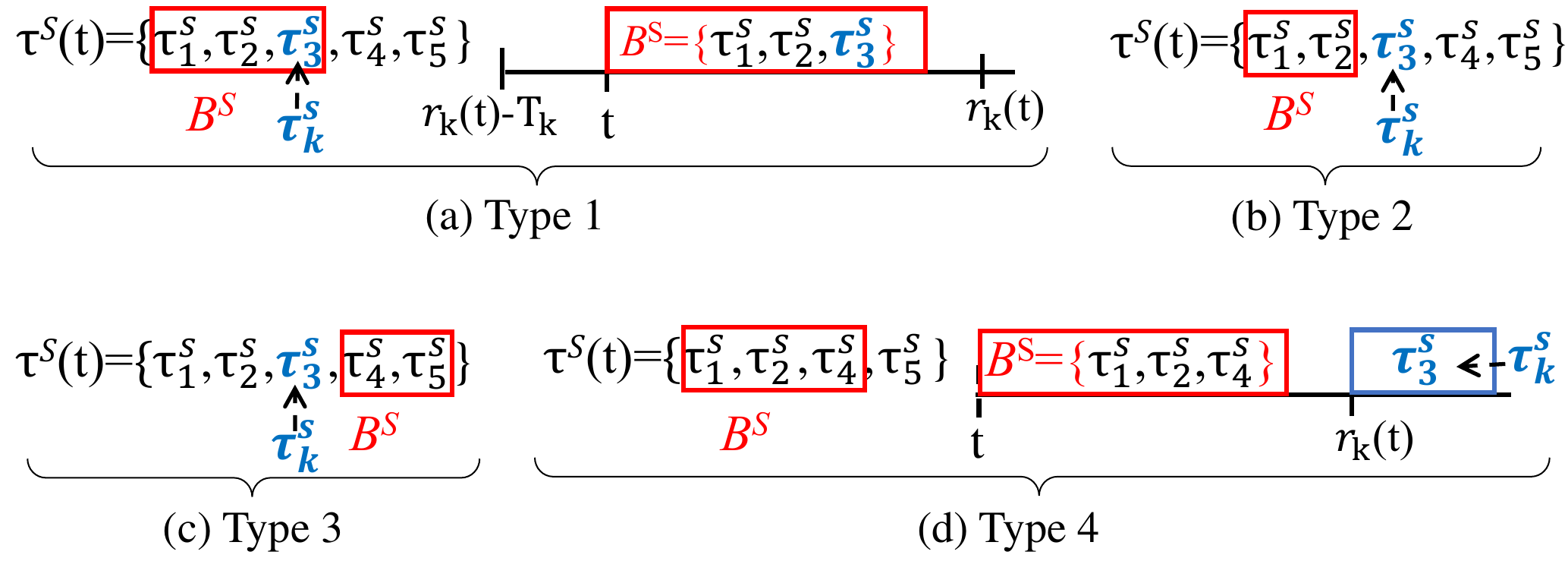}
    \caption{Four types of coarse subtask batching scenarios ($\mathcal{B}^S$) illustrating potential impacts on task executions under the \npfpstar{} framework, critical for schedulability analysis.}% \jlcmt{화살표 위치}}
    \label{fig:coarse_batch}
\end{figure}

Consider a task set $\tau$ deemed schedulable under \npfpc{} (per Lemma~\ref{lemma:offline}). 
If a \textsf{\small [C]}-enabled scheduler executes a coarse batch $\mathcal{B}^S$ at time $t$, its potential impact on any job $J_k$ (of $\tau_k^S \in \tau$) is categorized into four scenario types (illustrated in Fig.~\ref{fig:coarse_batch}). 
For this analysis, let $\tau_h^S$ be the highest-priority task in $\mathcal{B}^S$, $\tau^S(t)$ be the set of active coarse subtasks at $t$, and $r_k(t)$ be the earliest release of $\tau_k$ after $t$.
P1 and P2 (in conjunction with the batching property) guarantee that run-time coarse batching for each type preserves the schedulability ensured by Lemma~\ref{lemma:offline}, as follows. Recall that the batching property represents $C_{\mathcal{B}^S} \le \sum_{\tau_i\in\mathcal{B}^S} C_i^S$.

\vspace{5pt}
\begin{itemize}%[leftmargin=*]
    \item Type 1: $\tau_k^S\in\tau^S(t)$ and $\tau_k^S\in\mathcal{B}^S$ (e.g., Fig.~\ref{fig:coarse_batch}(a))
\end{itemize}
\begin{itemize}%[leftmargin=*]
    \item Type 2: $\tau_k^S\in \tau^S(t)$, $\tau_k^S\notin \mathcal{B}^S$, and $\tau_h^S\in\texttt{HP}(\tau_k^S)$ (e.g., Fig.~\ref{fig:coarse_batch}(b))
\end{itemize}

\noindent Under scenarios of Type 1 or 2, 
the higher-priority interference to tasks not in $\mathcal{B}^S$ will decrease (by the batching property), and the tasks in $\mathcal{B}^S$ finish their batching before their deadlines (by P2). 
\vspace{5pt}

\begin{itemize}%[leftmargin=*]
    \item Type 3: $\tau_k^S\in \tau^S(t)$, $\tau_k^S\notin \mathcal{B}^S$, and $\tau_h^S\in\texttt{LP}(\tau_k^S)$ (e.g., Fig.~\ref{fig:coarse_batch}(c))
    %$J_k$ may face additional blocking from $\mathcal{B}^S$ up to $C_{\mathcal{B}^S}$ (Fig.~\ref{fig:coarse_batch}(c)).
\end{itemize}

\noindent Type 3 scenarios are ruled out by P1. 
\vspace{5pt}

\begin{itemize}%[leftmargin=*]
    \item Type 4: $\tau_k^S\notin \tau^S(t)$ (e.g., Fig.~\ref{fig:coarse_batch}(d))
    %. Max interference on future $J_k$ from $\mathcal{B}^S$ is $\max(0, t+C_{\mathcal{B}^S}-r_k(t))$ instead of $\max(0, t+C_h^S-r_k(t))$ (Fig.~\ref{fig:coarse_batch}(d)).
\end{itemize}

\noindent Under Type 4 scenarios, $\tau_k$ is not affected by the batching execution (due to P2).
The impact of the batching on tasks in $\mathcal{B}^S$ and those not in $\mathcal{B}^S$ is equivalent to that observed in scenarios of Type 1 or 2.
\vspace{5pt}

Algo.~\ref{algo:coarse_batch} details the runtime logic for coarse batch assignment within any \npfpstar{} instantiation that supports \textsf{\small [C]}.
At each scheduling instant $t$, Algorithm~\ref{algo:coarse_batch} first checks if multiple coarse subtasks are active ($|\tau^S(t)| \ge 2$) (Line~1). 
If so, candidate batches $\mathcal{B}^S(k)$ (the $k$ highest-priority active coarse tasks) are defined (Line~2). 
The algorithm then seeks the largest $x \ge 2$ such that batch $\mathcal{B}^S(x)$ satisfies the runtime condition Eq.~\eqref{eq:coarse_batch_online} (Line~3). 
If such an $x$ is found (Line~4), this batch $\mathcal{B}^S(x)$ is executed (Line~5). 
Otherwise (if $|\tau^S(t)| < 2$ or no such $x$ is found), Line 8 executes the default: the single highest-priority active coarse subtask is run. 
This mechanism ensures that coarse batching is only employed when deemed safe by the runtime check. 
If no coarse subtasks are eligible for execution (either batched or individual), the overarching \npfpstar{} scheduler (Algo.~\ref{algo:npfp_star}) would then consider fine subtask execution based on its policy (e.g., Lines 7--12 of Algo.~\ref{algo:npfp_star}).

\vspace{5pt}
\begin{lemma}\label{lemma:coarse_batch_online_guarantee}
    Consider a task set $\tau$ deemed schedulable under \npfpc{} (per Lemma~\ref{lemma:offline}).
    Then, $\tau$ is also schedulable under \mbox{NPFP$^\textsf{[C]}$}.
\end{lemma}
\begin{IEEEproof}
    %\jlcmt{조금 rough함?}    
    (Active $\tau_k$) By P1, there is no scenario in which the batching lower-priority task(s) without including $\tau_k$ itself. %does not include an active $\tau_k$ but includes its lower-priority task(s), meaning that there are two cases: (i) batching does not include any lower-priority tasks, and (ii) batching includes lower-priority task(s).
    Thus, two cases arise: (i) the batch does not include any lower-priority tasks, or
    (ii) the batch includes one or more lower-priority tasks along with $\tau_k$.
    In case (i), due to the batching property, such batching does not increase the WCET of higher-priority interference, or of $\tau_k$ itself if it is included.
    In case (ii), the schedulability of $\tau_k$ is guaranteed by P2. 
    In both cases, P2 ensures that the batching execution does not affect any future job releases of $\tau_k$. 

    (Inactive $\tau_k)$ As explained in Type 4 scenarios, 
    P2 guarantees that batching execution does not interfere with future job releases of $\tau_k$.
\end{IEEEproof}

\setlength{\textfloatsep}{10pt}
\begin{algorithm} [t]
\caption{Coarse batch execution} 
\label{algo:coarse_batch}
\small
\raggedright
\begin{algorithmic}[1] % Added line numbers
    \IF {$|\tau^S(t)| \ge 2$} % Check if multiple coarse subtasks are active
        \STATE Let $\mathcal{B}^S(n)$ be the set of the $n$ highest-priority tasks in $\tau^S(t)$, $1\le n \le |\tau^S(t)|$.
        \STATE Find the largest $x$ ($2\le x\le|\tau^S(t)|$) s.t. $\mathcal{B}^S(x)$ satisfies Eq.~\eqref{eq:coarse_batch_online}. 
        %\st{and its
%WCET plus current time t is no later than the earliest future release of any coarse subtask.} \jlcmt{Eq.(5)와 and 뒤에 문구가 동일한거 아닌가?}
        \IF {such an $x$ exists}
            \STATE Execute tasks in $\mathcal{B}^S(x)$ as a batch, and \textit{return}.
        \ENDIF
    \ENDIF
    \STATE Execute the single highest-priority coarse active job, and \textit{return}. % Default action; \Delta t_next_coarse = time to earliest future coarse release
\end{algorithmic}
\normalsize
\end{algorithm}

\subsection{Fine batch assignment}
\label{subsec:fine_batch}

The fine batch assignment mechanism, detailed in Algo.~\ref{algo:fine_batch}, is invoked by an \npfpstar{} policy that supports batched fine subtask execution (\textsf{\small [F]}). 
This typically occurs when Line~11 of the generic Algo.~\ref{algo:npfp_star} is reached, specifically at a time $t$ when no coarse subtasks are active ($\tau^S(t) = \emptyset$) but active fine subtasks ($\tau^F(t) \neq \emptyset$) exist.
Upon invocation, the system first determines the workload (i.e., patch count $p^F \in \{\text{S, M, L}\}$ based on categories like small (S), medium (M), large (L) from patch splitting, cf. Fig.~\ref{fig:system_overview}~\ding{186}) for all $\tau_i^F \in \tau^F(t)$. 
These fine subtasks are then designated for execution using patch-level batching, potentially leveraging mechanisms like \hb{multi-level batch inference} (cf. Fig.~\ref{fig:system_overview}~\ding{188}).
A primary challenge in such fine subtask batching is the padding overhead: inputs with heterogeneous patch counts within a single batch are typically padded to match the maximum patch count (cf. Fig.~\ref{fig:system_overview}~\ding{187}), incurring latency costs dependent on batch composition. 
Furthermore, two critical timing constraints must be satisfied: (i) each fine subtask must meet its own deadline for its individual schedulability, and (ii) the entire sequence of fine subtask batches must complete execution before the earliest future release time of any coarse subtask to ensure coarse subtask schedulability is not compromised.

Finding a partition of the $|\tau^F(t)|$ active fine subtasks (with their determined workloads $w_i$) into batches $B_1^F, \ldots, B_x^F$ that minimizes the total predicted WCET $\sum_{y=1}^{x} C_{B_y^F}$ while satisfying constraints (i) and (ii) is computationally \jl{exponential} %intractable 
(potentially $O(2^{|\tau^F(t)|})$). 
Each batch WCET, $C_{B_y^F}$, is significantly influenced by padding overheads from mechanisms like \hb{multi-level batch inference}; for instance, if subtasks with different workloads (e.g., S, M, L) are batched, all are processed based on the maximum workload (e.g., L), and $C_{B_y^F}$ reflects the WCET for this padded configuration. 
Our goal is therefore to develop an efficient heuristic that reasonably reduces padding overhead and identifies a schedulable set of batches.
To efficiently partition active fine subtasks while reasonably reducing padding overhead, we propose the dynamic batch assignment (DBA) algorithm.

\setlength{\textfloatsep}{10pt}
\begin{algorithm}[t]
\caption{Fine batch execution}
\label{algo:fine_batch}
\small
\raggedright
\begin{algorithmic}[1]
    \IF {$|\tau^F(t)| \ge 2$} % Check for multiple active fine subtasks
        \STATE Let $\tau^F(t)'$ be the sequence of active fine subtasks from $\tau^F(t)$, sorted non-decreasingly by workload (e.g., patch counts).
        \STATE Determine an partition of $\tau^F(t)'$ into contiguous batches $B_1^F, \ldots, B_x^F$ that minimizes total predicted WCET $\sum_{y=1}^{x} C_{B_y^F}$ (per Lemma~\ref{lemma:dba_partition}).
        \STATE Execute the batch sequence $B_1^F, \ldots, B_x^F$, respecting fine subtask deadlines and the earliest future coarse subtask release, then \textit{return}.
        %\jlcmt{여기서 못돌리고 Line6으로 가는 경우는 없는지?}
    \ENDIF % Handles cases where $|\tau^F(t)| < 2$ (i.e., 0 or 1 fine subtask)        
        \STATE Execute the single highest-priority active fine subtask, if its WCET plus current time $t$ is no later than the earliest future release of any coarse subtask, then \textit{return}        
\end{algorithmic}
\normalsize
\end{algorithm}

\textbf{Principle.}
The core principle of DBA is to group fine subtasks with similar workloads, aiming to reduce WCET penalties from padding.
As detailed in Algo.~\ref{algo:fine_batch}, DBA first employs a heuristic step of sorting the $|\tau^F(t)|$ active fine subtasks by their workloads (e.g., patch counts) into a non-decreasing sequence $\tau^F(t)' = (w_1, \ldots, w_{|\tau^F(t)|})$ (Line~2).
Restricting batch formation to only \textit{contiguous} subtasks within this sorted sequence significantly reduces the search space for partitioning. Within this setting, DBA applies dynamic programming (DP) to determine a partition of $\tau^F(t)'$ into contiguous batches (Line~3). 
\hb{
This DP computation, guided by the recurrence in Eq.~\eqref{eq:dba_recurrence} and detailed in Lemma~\ref{lemma:dba_partition}, inherently incorporates system constraints. Specifically, if potential batches or sub-partitions violate either (i) individual fine subtask deadlines or (ii) the completion deadline set by the earliest next coarse subtask release, they are assigned an effective infinite cost and are thus excluded by the $\min$ operation in the recurrence.
}
%This DP computation, guided by the recurrence in Eq.~\eqref{eq:dba_recurrence} (detailed in Lemma~\ref{lemma:dba_partition}), inherently incorporates system constraints: potential batches or sub-partitions violating (i) individual fine subtask deadlines or (ii) the completion deadline before the next coarse subtask release are assigned an effective infinite cost, thus being excluded by the $\min$ operation in the recurrence.\jlcmt{문장 parsing이 잘 안됨. or or }
The DP stage, therefore, constructs a schedulable partition (if one exists satisfying the integrated constraints) for $\tau^F(t)'$ with an $O({|\tau^F(t)|}^2)$ time complexity. The resulting DP-derived partition is then processed by Line~4 of Algo.~\ref{algo:fine_batch} to dispatch an executable batch sequence.

\vspace{5pt}
\begin{lemma}
\label{lemma:dba_partition}
Let $\tau^F(t)' = (w_1, \ldots, w_N)$ be the sequence of $N=|\tau^F(t)|$ active fine subtask workloads, sorted non-decreasingly. The DP %dynamic programming (DP) 
stage of the DBA algorithm computes a contiguous batch partition of $\tau^F(t)'$ by evaluating the recurrence relation in Eq.~\eqref{eq:dba_recurrence}. Within this evaluation, terms corresponding to batches or sub-partitions that violate specified real-time constraints (\hb{i.e.,} (i) and (ii)) are effectively assigned infinite cost. The time complexity for this DP computation is $O(N^2)$. The WCET of a batch $C_{batch}(\cdot,\cdot)$, used in the recurrence, incorporates padding overheads and is assumed to provide its value in $O(1)$ time (e.g., via lookup from a WCET table).
\end{lemma}
\begin{IEEEproof}
The DP approach iteratively computes $\texttt{DBA}[k]$, representing the effective cost of a schedulable partition for the prefix $w_1, \ldots, w_k$, considering constraints integrated into the cost evaluation. With $\texttt{DBA}[0] = 0$ (zero cost and execution time), $\texttt{DBA}[k]$ for $k>0$ is found via:
\noindent
\begin{align}
\label{eq:dba_recurrence}
\texttt{DBA}[k] \;=\; \min_{1 \le j \le k} \left\{ \text{Cost}(j,k) \right\},
\end{align}
where $\text{Cost}(j,k) = \texttt{DBA}[j-1] \;+\; C_{batch}(w_k,\, k-j+1)$ if forming batch $(w_j, \ldots, w_k)$ and the path to it are schedulable (respecting constraints (i) and (ii)); otherwise, $\text{Cost}(j,k) = \infty$. Sorting ensures $w_k$ is the maximum workload in the last batch. Each $\texttt{DBA}[k]$ computation involves $O(k)$ evaluations of $\text{Cost}(j,k)$. Assuming $C_{batch}$ lookup and constraint checks for each term are efficient (e.g., $O(1)$ on average), computing $\texttt{DBA}[k]$ takes $O(k)$ time. Thus, the total time complexity to compute all $\texttt{DBA}$ values up to $N=|\tau^F(t)|$ is $\sum_{k=1}^{N} O(k) = O(N^2)$, with $O(N)$ space for the $\texttt{DBA}$ table.
\end{IEEEproof}
\vspace{5pt}

Table~\ref{tab:dp} exemplifies the DBA algorithm's computation (Eq.~\eqref{eq:dba_recurrence}, where terms can be $\infty$ if constraints are violated) for $|\tau^F(t)|=4$ fine subtasks with sorted workloads $w = \{\text{S}, \text{M}, \text{M}, \text{L}\}$ (S=1, M=2, L=3). Batch WCET $C_{batch}(w_k, \text{size})$ is $w_k$ if $\text{size}=1$, or $(w_k \times \text{size}) / 2$ if $\text{size} > 1$ (a 50\% reduction). The table details calculating $\texttt{DBA}[k]$ values. For instance, when computing $\texttt{DBA}[3]$ (for $w_3=\text{M}=2$), options include forming the last batch as $\{w_3\}$, $\{w_2, w_3\}$, or $\{w_1, w_2, w_3\}$. If an option (e.g., batch $\{w_1,w_2,w_3\}$ with $\texttt{DBA}[0]$) violates slack (constraint (ii)) or internal deadlines (constraint (i)), its $\text{Cost}(j,k)$ in Eq.~\eqref{eq:dba_recurrence} becomes $\infty$. $\texttt{DBA}[3]$ is the minimum of such costs. If the final $\texttt{DBA}[4]$ is finite (e.g., 5, yielding partition SM\,|\,ML), backtracking finds a schedulable sequence for all four tasks. Algo.~\ref{algo:fine_batch} (Line~4) then dispatches an executable prefix of this sequence (the full sequence if $\texttt{DBA}[4]$ allows). If $\texttt{DBA}[4]=\infty$, no full schedulable partition exists; a shorter valid prefix (from some $\texttt{DBA}[k<4]$) or no fine batches might be executed, always ensuring adherence to timing guarantees.

% 만약 두 컬럼 문서 레이아웃을 사용 중이라면, \begin{table*} ... \end{table*} 사용을 강력히 권장합니다.
% \begin{table*}[t!]
\begin{table}[t!]
\centering
\scriptsize % 글꼴 크기 유지
\caption{DP computation example (DBA[4]=5)}
\label{tab:dp} % 텍스트에서 참조하는 레이블과 동일하게 유지
\begin{tblr}{
  width = \linewidth, % 싱글 컬럼 문서의 경우. 두 컬럼 문서의 한 컬럼 너비는 \columnwidth
  colspec = {Q[l, font=\bfseries] *{10}{X[c]}}, % 첫 열: 왼쪽 정렬, 굵게. 나머지 10개 데이터 열: 자동 너비 조절, 가운데 정렬
  colsep = 3pt, % 열 간격 줄이기 (기본값보다 작게 설정)
  % 강조 표시된 열 (원래 테이블의 강조된 행에 해당)
  column{2} = {gray!30, font=\bfseries\scriptsize},
  column{4} = {gray!30, font=\bfseries\scriptsize}, % DBA[2] min
  column{7} = {gray!30, font=\bfseries\scriptsize}, % DBA[3] min
  column{9} = {gray!30, font=\bfseries\scriptsize}, % DBA[4] min
  % 가로선 (테이블의 상단 및 하단 테두리)
  hline{1,5} = {solid},
  % 세로선 (테이블의 왼쪽/오른쪽 테두리 및 헤더 열 구분선)
  vline{1,2,12} = {solid}, % 1 (헤더) + 10 (데이터) + 1 = 12
  % 굵은 세로선 (각 Step 그룹의 주요 경계)
  vline{3}  = {0.08em, solid}, % After Step 1 data group (orig: after 2nd data col)
  vline{5}  = {0.08em, solid}, % After Step 2 data group (orig: after 4th data col)
  vline{8}  = {0.08em, solid}, % After Step 3 data group (orig: after 7th data col)
  % 점선 세로선 (각 옵션 사이 구분)
  vline{4}  = {dotted},
  vline{6}  = {dotted},
  vline{7}  = {dotted},
  vline{9}  = {dotted},
  vline{10} = {dotted},
  vline{11} = {dotted}, % After 9th data col / Before last data col of Step 4
  rowsep = 1pt,
}
% 첫 번째 열: 원래 테이블의 헤더 (colspec에서 굵게 설정됨)
Step $k$                 & 1    & 2    & 2    & 3    & 3    & 3    & 4    & 4    & 4     & 4    \\
Last Grp. ($j..k$)       & 1--1 & 2--2 & 1--2 & 3--3 & 2--3 & 1--3 & 4--4 & 3--4 & 2--4  & 1--4 \\
Prev. $\texttt{DBA}[j-1]$ & 0    & 1    & 0    & 2    & 1    & 0    & 3    & 2    & 1     & 0    \\
Cost                     & 1    & 3    & 2    & 4    & 3    & 3    & 6    & 5    & 5.5   & 6    \\
\end{tblr}
\end{table}
% \end{table*} % 두 컬럼 문서에서 table*를 사용했다면 닫아줍니다.

\section{Evaluation}
\label{sec:evalution}

%\begin{figure}[t!]
%    \centering         
%    \subfloat[Server]{\includegraphics[width=0.33\linewidth]{figures/04scheduling/4x4_pooling_batch_latency_server_0429.pdf}} 
%    \subfloat[Orin]{\includegraphics[width=0.33\linewidth]{figures/04scheduling/4x4_pooling_batch_latency_0328.pdf}}
%    \subfloat[Overhead]{\includegraphics[width=0.33\linewidth]{figures/05evaluation/overhead_0506.pdf}}
%    \caption{...}
%    \label{fig:batch_latency}
%\end{figure}

%Our \sys{} were implemented using Python v3.10.12, PyTorch v2.1.0, and MMDetection v3.3.0~\cite{CWP19}.

\subsection{Experimental setup}
\label{sec:experimental_setup}

\textbf{Experimental environment.}
Experiments are conducted across three distinct hardware platforms: a server equipped with an NVIDIA A10 GPU, an NVIDIA Jetson Orin~\cite{Orin}, and a 1/10 scale AV featuring a Jetson TX2 board, which are utilized for the case study.
DINO, the transformer-based object detection model, operating at FP16 precision, is employed throughout these evaluations.
Its coarse stage, incorporating a ResNet-50 backbone~\cite{HZR16} and six encoder/decoder layers, is pre-trained on the COCO dataset~\cite{LMB14} and subsequently fine-tuned on KITTI~\cite{kitti12}.
Evaluations utilize the KITTI dataset, a standard autonomous driving benchmark comprising 8,110 urban RGB images from 21 sequences and covering eight classes (car, van, truck, pedestrian, person sitting, cyclist, tram, misc).
This dataset is equally divided, with one half allocated for model training and fine-tuning, and the other reserved for assessing the performance of the \sys{} pipeline and its scheduler.

\begin{table}[t]
\centering
\caption{Execution time measurement (ms) for components}
\label{tab:WCET}
\renewcommand{\arraystretch}{0.7}
\resizebox{1\columnwidth}{!}{
%\normalsize
\begin{tblr}{
  colspec = {Q[m,c] | c | ccc | cccc},
  row{4,6} = {gray!30, font=\bfseries}, % WCET 행에 볼드체(font=\bfseries) 추가
  hlines,
  rowsep = 0pt,
  cell{1}{1} = {r=2}{m},
  cell{1}{3} = {c=3}{c},
  cell{1}{6} = {c=4}{c},
  cell{3}{1} = {r=2}{m},
  cell{5}{1} = {r=2}{m},
}
Platform & Component   & $C_i^S$      &                 &          & $C_i^F$         &             &             &             \\
         & Time(ms) & $c_i^{ps}$   & $c_i^{at}(p_i^S)$ & $c_i^{dt}$ & $c_i^{sps}$     & $c_i^{at}(S)$ & $c_i^{at}(M)$ & $c_i^{at}(L)$ \\
Server   & Mean     & 20         & 45             & 0.2      & 2             & 44          & 52          & 55          \\
         & WCET     & 30         & 49             & 0.3      & 3             & 46          & 55          & 58          \\
Orin     & Mean     & 59         & 68              & 0.5      & 5             & 65          & 80          & 85          \\
         & WCET     & 70         & 69              & 0.7      & 8             & 78          & 96          & 107         \\
\end{tblr}
}
%\vspace{-0.5cm}
\end{table}

\begin{figure}[t!]
    \centering
    \includegraphics[width=0.95\linewidth]{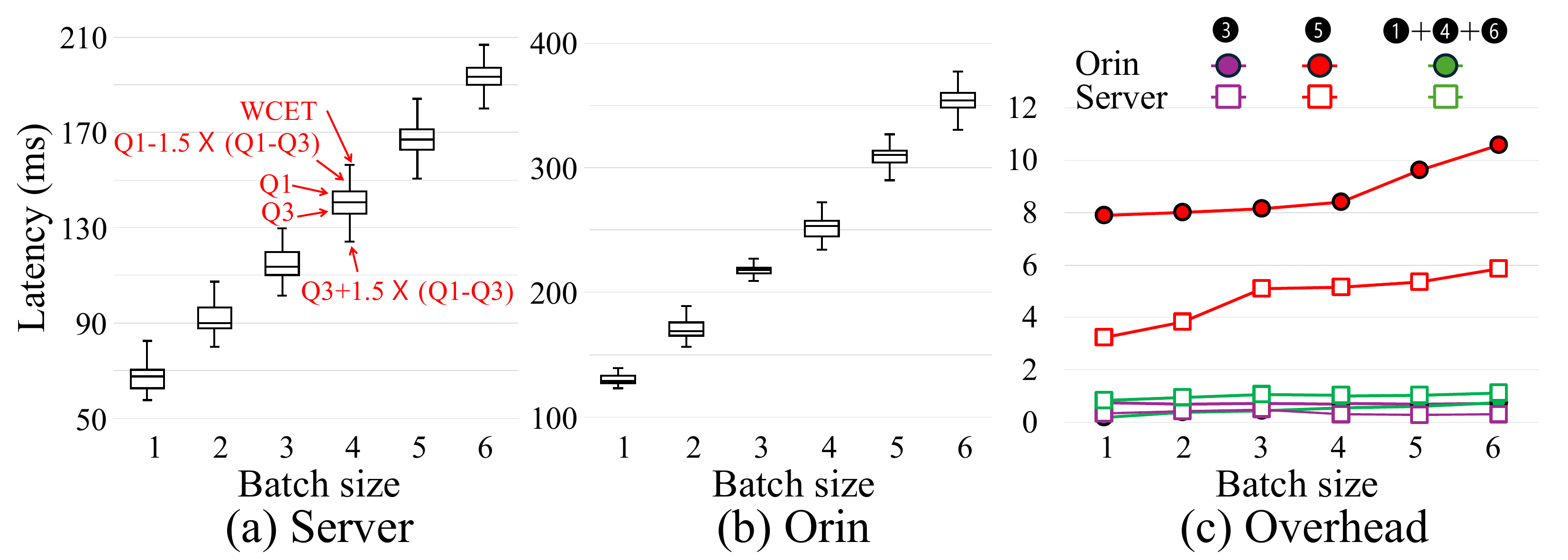}
    \caption{Latency of coarse batch execution by size on (a) server and (b) Orin, and (c) average runtime overhead of \textsf{CF-DETR}'s core components by batch size.}
    \label{fig:batch_latency}
\end{figure}

\textbf{Execution time profiling and run-time overhead.}
WCETs of individual components for coarse subtasks ($C_i^S$) and fine subtasks ($C_i^F$) are determined via 1000 measurements each on both the server and the Jetson Orin platform, with detailed results presented in Table~\ref{tab:WCET}. 
%For instance, on the server, key coarse subtask components ($c^{ps}_i$, $c^{at}_i(p^S_i)$, and $c^{dt}_i$) exhibit measured times of 30\,ms, 49\,ms, and 0.3\,ms, respectively. 
%Correspondingly, components of the fine subtask ($c^{sps}_i$, $c^{at}_i(\text{L})$, $c^{at}_i(\text{M})$, and $c^{at}_i(\text{H})$) record times such as 3\,ms, 46\,ms, 55\,ms, and 58\,ms.
Figs.~\ref{fig:batch_latency}(a) and (b) illustrate the latency distributions for coarse batch sizes ($|\mathcal{B}^S|$) ranging from one (no batch) to six on these platforms. 
Furthermore, Fig.~\ref{fig:batch_latency}(c) depicts the average run-time overhead of core operational components: hardness identification (cf. $c^{dt}_i$, \ding{184} in Fig.~\ref{fig:system_overview}), selective patch split (cf. $c^{sps}_i$, \ding{186} in Fig.~\ref{fig:system_overview}), and scheduling overhead (cf. Algo.~\ref{algo:npfp_star}, \ding{182}+\ding{185}+\ding{187} in Fig.~\ref{fig:system_overview}) for varying batch sizes. 
Specifically, the overhead for hardness identification ($c^{dt}_i$) and scheduling is minimal (under 1\,ms). 
Selective patch split ($c^{sps}_i$) incurs approximately 8--11\,ms, depending on the batch size, which is a reasonably small addition compared to the inference time ($c^{at}_i(p)$).

\begin{figure}[t]
    \centering
    \subfloat[Accuracy on server (A10 GPU)]{\label{fig:main_eval_server}
        \includegraphics[width=0.9\linewidth]{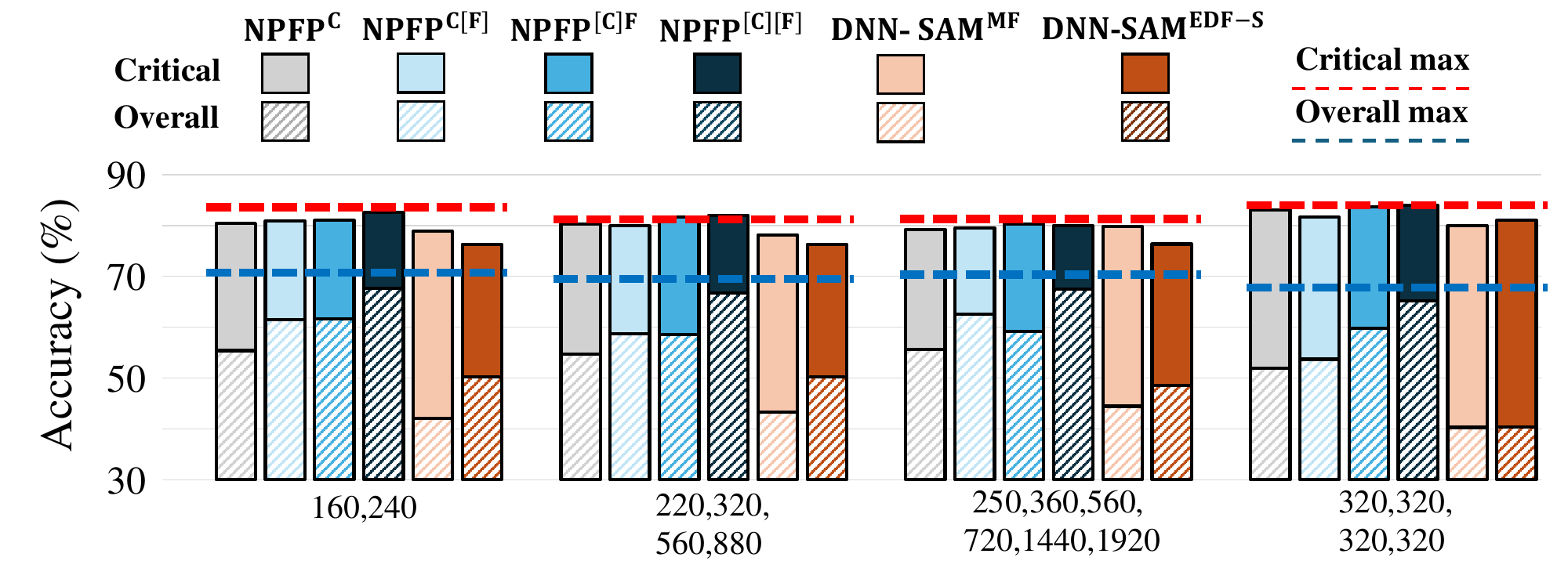}
    }
    \vspace{0.01cm}
    \subfloat[Accuracy on Jetson Orin]{\label{fig:main_eval_orin}
        \includegraphics[width=0.9\linewidth]{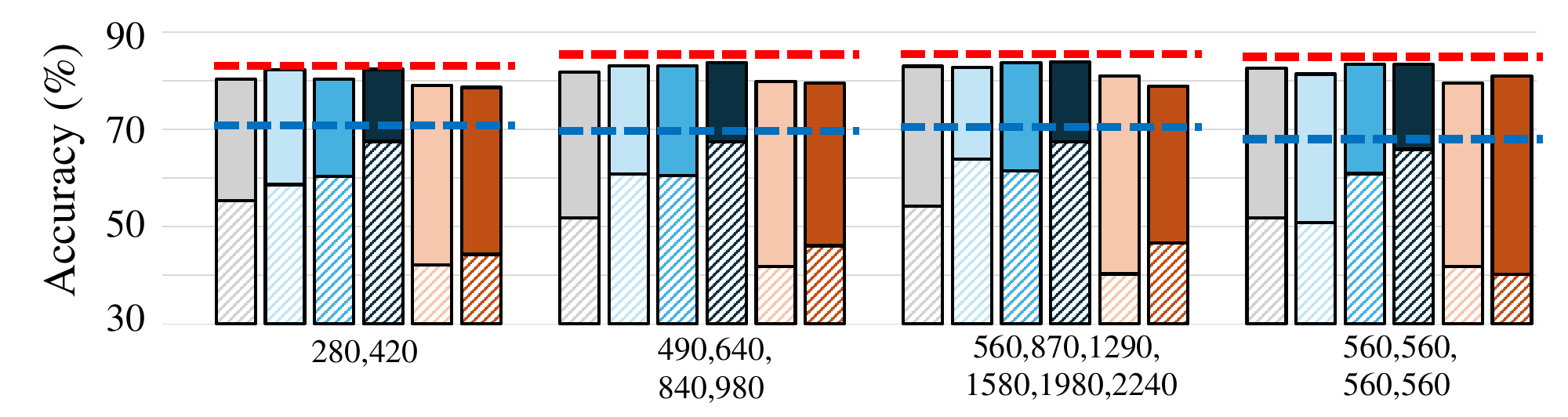}
    }

    \caption{Critical and overall mAP (\%): \textsf{CF-DETR} (under \npfpstar{} variants) vs. DNN-SAM, on (a) server and (b) Jetson Orin across various task sets.}
    \label{fig:main_eval_combined}
\end{figure}

\subsection{Experimental results}
\label{sec:experimental_result}

We evaluate the following four \sys{} scheduling algorithms \npfpstar{} on their effectiveness in improving accuracy (R2) while meeting real-time constraints (R1):
\begin{itemize}%[leftmargin=*]
    \item \npfpc{} executing only individual coarse subtasks without batching,
    \item \npfpcfb{} with individual coarse subtasks and batched fine subtasks,
    \item \npfpcbf{} with batched coarse subtasks and individual fine subtasks, and
    \item \npfpcbfb{} where both coarse and fine subtasks are batched.
\end{itemize}

\noindent
We compare these against two DNN-SAM scheduling algorithms, an existing approach for R1/R2 satisfaction in multi-MOD tasks:
\begin{itemize}%[leftmargin=*]
    \item \mbox{DNN-SAM$^\textsf{MF}$} (MandFirst) that favors mandatory subtask frames per second (FPS) by always prioritizing it over the optional subtask, and
    \item \mbox{DNN-SAM$^\textsf{EDF-S}$} (EDF-slack) that favors overall accuracy by adaptively prioritizing the optional subtask via runtime slack reclamation.
\end{itemize}
All six evaluated algorithms uniformly employ DINO~\cite{ZLL22} as the detector. 
For all evaluations, task sets satisfy offline schedulability tests: \sys{} algorithms adhere to Lemma~\ref{lemma:offline} under rate monotonic (RM)~\cite{rm} priority assignment, while DNN-SAM algorithms follow Theorem 1 in~\cite{KCK22} using earliest deadline first (EDF)~\cite{LCLJ73}. 
Detection accuracy is measured by mean average precision (mAP)~\cite{LMB14}; ``critical accuracy'' is for critical objects (area $>$ 16384 pixels), and ``overall accuracy'' for all objects (including critical ones). ``Critical max'' and ``overall max'' denote DINO's maximum achievable accuracy (without schedulability guarantees). Task response times are measured in FPS.

Fig.~\ref{fig:main_eval_combined} presents critical and overall accuracy results on two platforms. 
Experiments use task sets of two, four, or six tasks with varying periods (detailed on the x-axis of Fig.~\ref{fig:main_eval_combined}), all satisfying Lemma~\ref{lemma:offline}, ensuring guaranteed coarse subtask execution while fine subtask processing remains opportunistic. 
For critical accuracy, all \sys{} variants achieve near-maximum performance, outperforming DNN-SAM algorithms; note that critical accuracies can vary slightly among \sys{} variants as fine subtasks may also detect critical objects missed by their coarse counterparts. 
This superiority stems from \sys{}'s low-latency coarse patch attention effectively detecting large, dispersed critical objects while minimizing focus on irrelevant background. In contrast, the model-agnostic DNN-SAM processes a single, rescaled, and potentially oversized region for all critical objects (including background), which can reduce critical accuracy. 
Regarding overall accuracy, DNN-SAM algorithms show a significant performance drop due to a lack of batch support and potentially requiring two full inferences if optional subtasks are invoked, without resource-effective adaptation. 
Conversely, \sys{} variants, especially with multi-level batched execution, maintain higher overall accuracy.

\begin{figure}[t]
    \centering
    \includegraphics[width=0.9\linewidth]{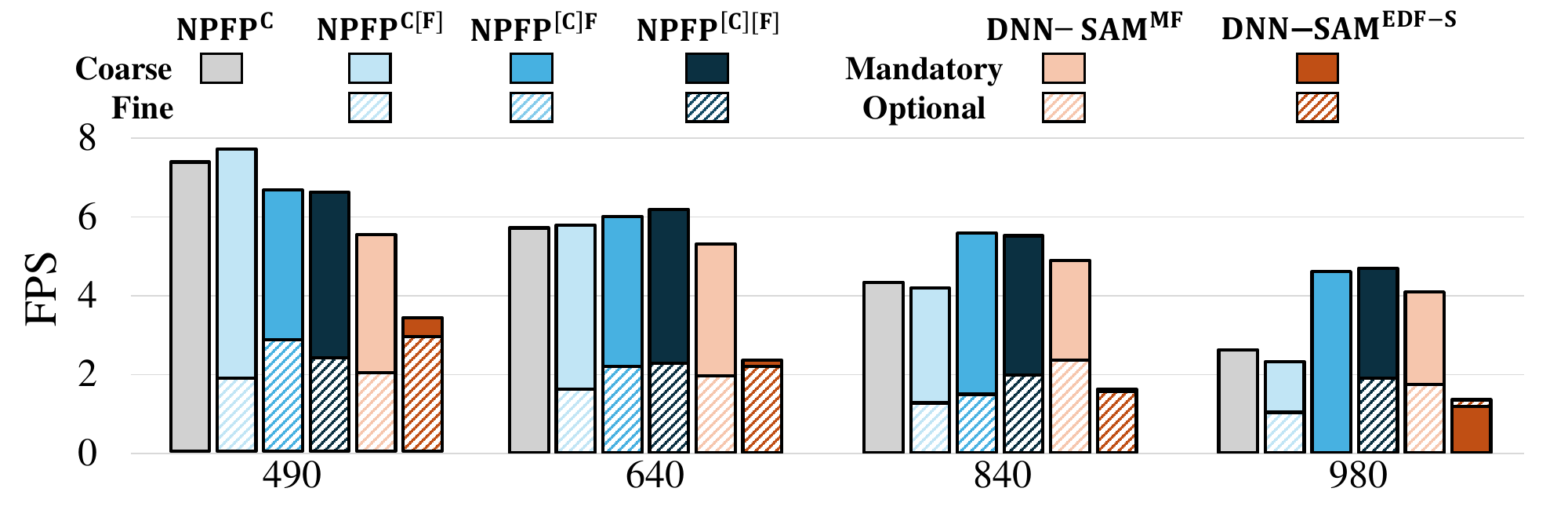}
    \caption{Per-task FPS: \textsf{CF-DETR} (under \npfpstar{} variants) vs. DNN-SAM on Jetson Orin for four-task sets (490,640,840,980).}
    \label{fig:main_eval_combined_fps}
\end{figure}

Fig.~\ref{fig:main_eval_combined_fps} shows per-task FPS results for representative task sets on the Jetson Orin platform, with tasks arranged by decreasing RM priority. Under \sys{}'s policies, coarse subtasks are strictly prioritized, ensuring their execution is unimpeded by fine subtasks; this guarantees %R1 
responsiveness for critical detections while enabling efficient fine-detail processing. Consequently, policies lacking efficient coarse subtask batching (e.g., \npfpc{} and \npfpcfb{}), or those not prioritizing FPS like \mbox{DNN-SAM$^\textsf{EDF-S}$}, exhibit a marked FPS reduction for lower-priority tasks. 
\hb{Note that the FPS difference among tasks under NPFP$^{\textsf{C}*}$ stems from fixed-priority scheduling, but is mitigated through coarse subtask batching in NPFP$^{\textsf{[C]}*}$.}

%\textsf{\small DNN-SAM}, in particular, shows significantly lower FPS due to its potential need for two individual inferences per task. 
%\textsf{\small DNN-SAM}, exhibit a marked FPS reduction for lower-priority tasks; \textsf{\small DNN-SAM} is particularly affected due to its potential need for two full individual inferences per task instance. 
%In contrast, \sys{} policies that incorporate batch execution for coarse subtasks, such as \npfpcbf{} and \npfpcbfb{}, sustain more consistent and higher FPS across all tasks. 
%These findings demonstrate that effective batching approaches can mitigate task starvation and maintain high throughput. 
%Furthermore, 
%token-level region partitioning, which effectively minimizes focus on irrelevant background areas, whereas \textsf{\small DNN-SAM}'s safety-critical regions, identified via rough object-based cropping, can include such areas. 
%\hb{CF-DETR은 object단위로 잘 하지만 DNN-SAM은 모든 critical object가 있는 큰 영역을 single region을 crop한다.}
%This is significantly aided by multi-task attention for fine subtasks, a DETR-specific batching method discussed in Sec.~\ref{sec:system_design} that substantially reduces latency by efficiently managing execution times and resources.
%; all sets satisfied Lemma~\ref{lemma:offline}, which guarantees individual coarse subtask execution, though fine subtask execution remains opportunistic.
%both the server (A10 GPU) and the NVIDIA Jetson Orin platform. 

\subsection{Case study: emergency braking}
\label{subsec:case_study}

To validate the real-world practicality of \sys{}, we conducted an emergency braking case study on a 1/10 scale AV equipped with a Jetson TX2 board. 
The vehicle, featuring two cameras (front and rear MOD tasks with periods of 1600\,ms and 2400\,ms, respectively), LiDAR, an IMU sensor, and multiple actuators, operated at a speed of 0.5--0.6\,m/s and was programmed to stop upon detecting an object within 1.5\,m. 
Execution times for \sys{} components on the TX2 platform are detailed in Table~\ref{tab:TX2_WCET}. 
We compared \sys{} under the \npfpcbfb{} policy against the DINO baseline, performing 10 trials with each approach.

\begin{figure}[t]
    \centering
    \includegraphics[width=0.8\linewidth]{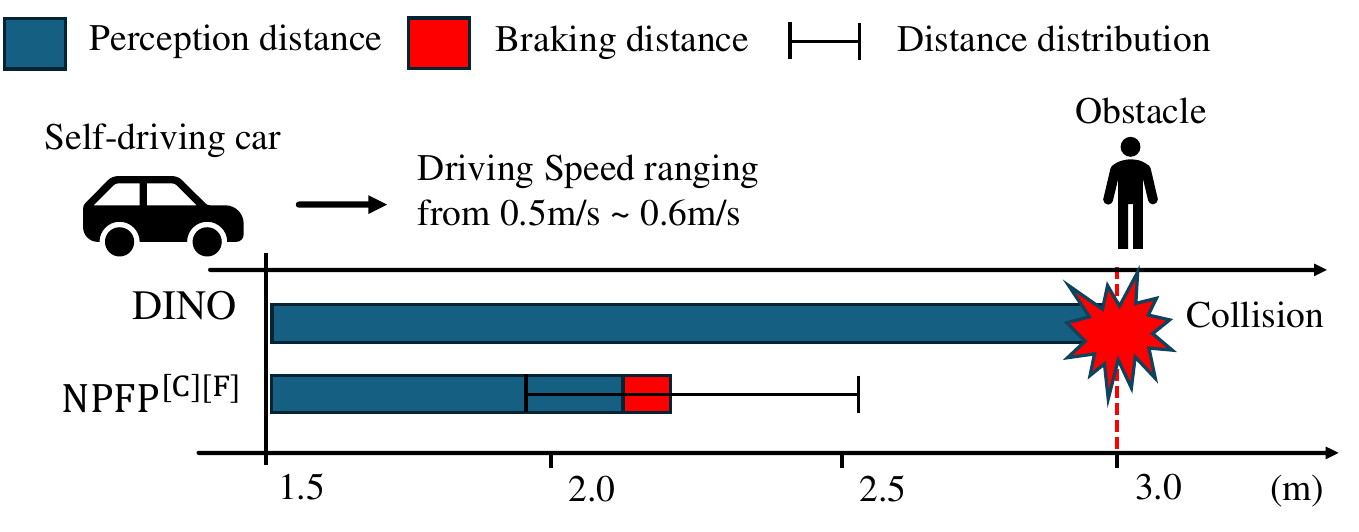}
    \caption{Emergency braking: \mbox{NPFP$^\textsf{[C][F]}$} ensures safe stopping for critical objects, while achieving overall accuracy comparable to DINO (cf. Fig.~\ref{fig:main_eval_combined}).} 
    \label{fig:xycar}
\end{figure}

The results, summarized in Fig.~\ref{fig:xycar}, highlight \sys{}'s superior performance in ensuring safety. 
In all trials, DINO failed to detect the obstacle and stop the car before a collision, primarily due to the high perception latency incurred by its reliance on fine subtask processing for detection. 
In stark contrast, \sys{} using the \npfpcbfb{} policy successfully avoided collisions in every trial. 
%This success is attributed to its strategy of prioritizing the rapid execution of low-latency coarse subtasks, which ensures timely detection of critical objects. 
\sys{} consistently brought the vehicle to a stop at a safe distance of 0.47--1.01\,m from obstacles, with an average stopping distance of 0.88\,m. 
This demonstrates the practical effectiveness of \sys{} in safety-critical scenarios, achieving reliable obstacle avoidance while, as noted in the caption of Fig.~\ref{fig:xycar}, maintaining high overall accuracy comparable to that of DINO (cf. Fig.~\ref{fig:main_eval_combined}).

\section{Related Work}
\label{sec:related_work}

Transformers, demonstrating superior accuracy over traditional CNNs/RNNs, have transitioned from NLP success to broad adoption in computer vision tasks like image classification~\cite{DBK20} and object detection~\cite{CMS20}. 
Their substantial computational cost and latency, however, have spurred research into lightweight architectures~\cite{YVA22, BFD22}, such as A-ViT~\cite{YVA22} with token-halting and ToMe~\cite{BFD22} with token merging. 
While effective in reducing average computational demands, these methods often lack reliable focus on safety-critical objects and regions required by safety-critical systems. 
Similarly, CF-ViT~\cite{CLL23} introduced a coarse-to-fine strategy for classification by leveraging token size versus accuracy, but it does not address safety considerations or detection tasks.

Research in RT-DNN scheduling~\cite{KCK22, KLC22} has targeted concurrent timely execution and accuracy satisfaction; approaches include DNN-SAM~\cite{KCK22} prioritizing critical regions and RT-MOT~\cite{KLC22} dynamically allocating resources for tracking, though often with suboptimal GPU utilization due to limited batching. 
Conversely, another line of work improved GPU utilization via batching~\cite{SXM22, ZBL18, XYK19} for responsiveness (e.g., in single multi-object detection tasks~\cite{SXM22}, S$^3$DNN~\cite{ZBL18}, DART~\cite{XYK19}), but typically overlooked hard real-time constraints. 
More recently, Batch-MOT~\cite{KLH24} aimed to combine batching with strict timing for tracking; however, it lacked efficient batch execution tailored for detection Transformers and a focus on safety-critical regions, prioritizing overall accuracy instead.
\begin{table}[t]
\centering
\caption{Execution time measurement (ms) for components}
\label{tab:TX2_WCET}
\renewcommand{\arraystretch}{0.7}
\resizebox{1\columnwidth}{!}{
%\normalsize
\begin{tblr}{
  colspec = {c|c | ccc | cccc},
  row{4,6} = {gray!30, font=\bfseries}, % WCET 행에 볼드체(font=\bfseries) 추가
  hlines,
  rowsep = 0pt,
  cell{1}{1} = {r=2}{m},
  cell{1}{3} = {c=3}{c},
  cell{1}{6} = {c=4}{c},
  cell{3}{1} = {r=2}{m},
  cell{5}{1} = {r=2}{m},
}
Platform & Component& $C_i^S$     &                   &            & $C_i^F$       &               &               &               \\
         & Time(ms) & $c_i^{ps}$  & $c_i^{at}(p_i^S)$ & $c_i^{dt}$ & $c_i^{sps}$     & $c_i^{at}(S)$ & $c_i^{at}(M)$ & $c_i^{at}(L)$ \\
TX2      & Mean     &    349      &      288          &    1.1     &      32       &    962        &    1207       &  1441         \\
         & WCET     &    408      &      368          &    1.5     &      38       &    1147       &    1245       &  1478         \\
\end{tblr}
}
%\vspace{-0.5cm}
\end{table}

\section{Conclusion}
In this paper, we proposed \sys{} to overcome the limitations of existing generic RT-DNN approaches in concurrently executing multiple detection tasks for AV systems, addressing the crucial R1 \& R2 balance challenge posed by DETR's latency-accuracy trade-off. 
\sys{}'s novelty lies in being the first to adapt the coarse-to-fine paradigm, previously explored in classification, to the complexities of object detection, thereby creating an architectural foundation (i.e., A1--A3) to effectively navigate this trade-off.
Building upon this, the \npfpstar{} scheduler (i.e., A4) is specifically designed to achieve the overarching goals of R1 while achieving high accuracy requirement R2. %(R2).
%, by prioritizing A1-driven coarse subtasks, and R2, by leveraging the selective (A2) and batched (A3) fine-grained processing. 
Extensive evaluations confirmed that \sys{} consistently meets critical deadlines, validated by superior FPS and successful AV emergency braking, while significantly outperforming baselines in both critical and overall detection accuracy. 

\bibliographystyle{IEEEtran}
\bibliography{main}

\end{document}